# Controlling the morphology and outgrowth
# of nerve and neuroglial cells:
# The effect of surface topography

C. Simitzi, A. Ranella and E. Stratakis

*Institute of Electronic Structure and Laser (IESL), Foundation for Research and Technology-Hellas (FORTH), Heraklion, 71003, Greece*

**Abstract**

Unlike other tissue types, like epithelial tissue, which consist of cells with a much more homogeneous structure and function, the nervous tissue spans in a complex multilayer environment whose topographical features display a large spectrum of morphologies and size scales. Traditional cell cultures, which are based on two-dimensional cell-adhesive culture dishes or coverslips, are lacking topographical cues and mainly simulate the biochemical microenvironment of the cells. With the emergence of micro- and nano-fabrication techniques new types of cell culture platforms are developed, where the effect of various topographical cues on cellular morphology, proliferation and differentiation, can be studied. Different approaches (regarding the material, fabrication technique, topographical characteristics, etc.) have been implemented. The present review paper aims at reviewing the existing body of literature on the use of artificial micro- and nano-topographical features to control neuronal and neuroglial cells' morphology, outgrowth  and neural network topology. The cell responses–from phenomenology to investigation of the underlying mechanisms- on the different topographies, including both deterministic and random ones, are summarized.

**Statement of significance**

With the aid of micro-and nanofabrication techniques, new types of cell culture platforms are developed and the effect of surface topography on the cells has been studied. The present review article aims at reviewing the existing body of literature reporting on the use of various topographies to study and control the morphology and functions of cells from nervous tissue, i.e. the neuronal and the neuroglial cells. The cell responses–from phenomenology to investigation of the underlying mechanisms- on the different topographies, including both deterministic and random ones, are summarized.

**Table of contents**

## 1. Introduction
## 1.1 The *in vitro* study of the effect of topography on nerve cells

The role of soluble (bio)chemical signals in cell shape, cell adhesion, differentiation and axon guidance, is well established [1]. In addition to the biochemical signals, there is increasing evidence that the physical parameters (e.g. topography and stiffness) of the complex extracellular milieu, constituted by the extracellular matrix (ECM) components and the surrounding cells, are also important [2]. Recent *in vitro* studies suggest that many neuronal and neuroglial cells respond to stiffness and to other mechanical cues during development (reviewed in [3]). Topography, which can be described as the arrangement of the spatial and structural features (e.g. geometrical architecture, discontinuities motif, contours, etc.) of the extracellular environment, has been also strongly correlated with specific functional characteristics of the cells and the tissues at both physiological (e.g. during development) and pathological states (e.g. wound healing) of nervous tissue [2,4].

The importance of ECM architecture in cells and tissue organization is apparent already from the early developmental stages. Embryonic cells produce their own extracellular scaffolds by secreting many types of molecules in the surrounding space, following a well defined program of differentiation [5,6]. The different spatial organization of these secreted molecules gives rise to a great variety of natural scaffolds where cells continue to proliferate and organize themselves in order to build up tissues and accomplish all their natural functions [7]. The role of ECM organization and distribution on directing neural crest migration has been early understood [8,9]. During glioma progression, i.e. a brain cancer, individual cancer cells have the tendency to migrate along myelinated fibers of white matter tracts [10].

The role of topography on cellular outgrowth *in vitro* has been addressed very early; already in 1914 *R. G. Harrison* cultured embryonic frog spinal neurons in a meshwork of spider web filaments and observed that such cells preferentially extended along the solid support of the filaments. Some years later, in 1934, *P. Weiss* made similar observations with embryonic chicken spinal neurons on grooves generated by brushing clotting blood and established the term "contact guidance", in an attempt to describe the tendency of the cells to orient themselves along anisotropic topographical features of the surface (such as fibers or ridges) [2,11]. The observations that certain physical properties of the substrate can *in vitro* influence cellular outgrowth and functions opened a new promising research field. However, many of these early experimental situations in which contact guidance was demonstrated were quite complex. Indeed, it was difficult to discriminate between the effects resulting from the chemical cues and those resulting from the topographical ones. More specifically, the substrates used, i.e. plasma clots, fish scales, and various grooved surfaces, were anisotropic not only in shape but also in chemistry [12].

Accordingly, there was a need to carefully study the effect of topographical cues *vis-a-vis (*bio)chemical cues in a reproducible way. With the emergence of micro- and nanofabrication techniques, a plethora of approaches to engineering or tailoring surfaces in a controllable manner are now available and specific topographical patterns, at micro and sub-micron scale can be designed and fabricated at a plethora of different materials [13–16]. Patterning of surfaces has triggered the development of new types of cell culture platforms, where the effect of topographical cues on cellular responses can be investigated and/or manipulated, depending on the field of interest.

## 1.2 Neurons and neuroglial cells: the basic components of the nervous tissue

Unlike other tissues, for example the epithelial tissue, where cells exhibit simple shapes, the nervous tissue is a complex three-dimensional environment whose topographical features span a large spectrum of morphologies and size scales. Nerve cells are the functional units of the nervous tissue which are responsible for transmitting information from and towards the environment through electrical and chemical signals. Different neuronal cell subtypes exist in the central and peripheral nervous tissue, namely the pyramidal cells and Purkinje cells in the central nervous system (CNS) and the sensory, motor and sympathetic neurons in the peripheral nervous system (PNS). Nerve cell shape exhibits a highly polarized pattern, comprising two types of neurite extensions from the cell body, namely the dendrites and the axon. This polarization, termed as neuronal polarity, establishes the signal transmission which underlies neural function. Cell bodies vary significantly in size among the different cell types; however, in the vertebrate nervous system, they typically exhibit diameters in the range of 10–50 μm, whereas axons and dendrites have diameters in the micrometer range (typically 0.2–3 μm) [17].

The growth cone, the highly motile leading edge of axon, navigates along specific pathways to reach the correct target, by recognizing and translating an ensemble of bio(chemical) and physical cues, which are mediated either by diffusion or by contact. Growth cone uses its cytoskeletal elements, including actin filaments and microtubules, to move forward in a process that involves the aid of a plethora of regulator proteins, such as the motor protein myosin II [18]. In this way, the growth cone is advanced and the axon is elongated (Axon outgrowth). Furthermore, growth cone uses the elements of the navigation system, including signal transduction molecules (e.g. the Rho family of GTPases that control many downstream signalling pathways) which translate environmental attractive or repulsive guidance cues into localized cytoskeletal remodelling (Axon guidance) [19]. Axon guidance can share similar mechanisms with the axon branching, i.e. the formation of axonal branches arisen by splitting of the terminal growth cone, which is another important axonal process in CNS development [20,21]. Extending axons may also grow in tight bundles (fasciculation). Fasciculation is driven by a balance of attractive and repulsive forces on the axons relative to their surrounding environment and by cues provided by other axons [22]. All these neuronal processes briefly described above, including the neuronal polarity establishment, axonal outgrowth, guidance and fasciculation, shape the neuronal morphology and are critical for the formation of the intricate neural networks of the functional neural tissue during development and regeneration.

Neuroglial cells present the other main cell type of nervous tissue and provide a rich and supportive environment for neurite outgrowth during development and nerve regeneration. Astrocytes, oligodendrocytes, and microglial cells are three types of glial cells in the mature CNS and Schwann cells are the cells that elaborate myelin in the PNS. Owing to their capability to release neurotrophic factors, to express cell surface ligands and synthesize extracellular matrix (ECM) but also to their oriented shape and structural organization, neuroglial cells present both molecular and topographical guidance stimuli for the development and outgrowth of neurons. A characteristic example includes radial glia, which are organized into regular arrays spanning the walls of the developing brain and guide the migrating embryonic neurons [23].

## 1.3 3D micro/nano surface texturing of biomaterials as a means of manipulation of neuron cell fate

Numerous techniques are available for the development of patterned biomaterial surfaces, ranging from simple manual scratching to more controlled fabrication methods [24]. Examples of techniques to engineer surfaces in a controllable manner include photolithography, microcontact printing, microfluidic patterning and electrospinning [13,15,16,25]. Using these techniques, 3D topographical features of tailored geometry, roughness and orientation, complemented by the desired spatial resolution at micron- and submicron-scales, can be realized on material surfaces of different chemical and mechanical properties [25,26]. The main characteristics of the 3D micro/submicron surface texturing techniques of biomaterials are listed in the Table 1 [16,25,27–33]. With the aid of these techniques mainly the well-defined geometries, also coined as deterministic can be fabricated.

Furthermore, techniques and strategies to fabricate more random surface morphologies at nanoscale have been implemented. Usually, such substrates are fabricated via chemical or physical etching and surface roughness can be controlled by variable exposure time to chemicals or physical agents, like plasma. Another approach for the development of surfaces exhibiting random roughness is the fabrication of monodispersed colloids (e.g. the fabrication of monodisperse silica particles via the Stoeber process). There are few studies using nanoporous gold, which is typically produced by de-alloying of silver-gold alloys, i.e. by selective dissolution of silver from a silver-rich gold alloy [34].

Table 1a: Main characteristics of 3D micro/submicron surface texturing techniques of biomaterials

| Technique | Principle | Material | Pattern Type | Advantages | Disadvantages | Ref. |
|---|---|---|---|---|---|---|
| <u>Photo-structuring:</u> Photo-lithography | The transfer of a user-generated geometrical pattern onto a material through the selective exposure of a light sensitive polymer | Usually silicon | Usually: alternating grooves and ridges | -The basis for other microfabrication techniques, such as electron beam, X-ray lithography, etc. -The underlying processes for all microstructuring processes in microelectronics | - Not easily applied to curved surface - Multiple step technique - Clean-room facilities & expensive equipment required | [16] |
| <u>Photo-structuring:</u> Ultrashort-pulsed Laser ablation | A pulsed laser beam of high intensity at short timescale impinges on substrate. Material is vapourized or melted and ejected from surface. Pattern can emerge by x-y movement of the sample, the beam or a combination of both | Broad range of materials (ceramics, metals and polymers) | Complex geometries including 3D shapes or structures with varying wall shapes and etch depths, and various aspect ratios on the same substrate | -Limited size of the affected volume -High fabrication rate -Non-contact interaction -Reproducibility | -Specific feature geometries not possible -Limited control on surface chemistry | [31] |
| <u>Photo-structuring:</u> LIGA | A hybrid fabrication technique consisting of sequentially 1.lithography, 2.electroplating and 3. molding | Wide variety: polymers, metals, alloys and ceramics | High aspect ratio microstructures | -Large structural height and sidewall properties - Thickness ranging from 100-1000 µm - High spatial resolution - High aspect ratios | - High cost in some types (i.e. X-ray LIGA) - Slow process - Complicated process - Research to production transition is difficult | [29] |
| E-beam lithography (EBL) | A computer-controlled electron gun scans an electron beam across the electron-sensitive resist coated on the surface. Both positive and negative type resists are available | Usually silicon or glass and an electron-sensitive resist (e.g. PMMA) | High precision patterns at nanoscale | -No need for mask. -Computer-controlled patterning | -High cost -Time consuming -Small surface coverage -Negative resists result in lower feature resolution | [25] |
| Colloidal lithography | Use of colloidal crystals as masks for etching and deposition for the fabrication of various nanostructures on planar and non-planar substrates | Wide range of materials (e.g. polymers and metals). | Patterns vary from irregular or random to pseudo-regular | -Simple and accessible -High spatial resolution -Easier to pattern larger areas than with EBL | Specific feature geometries not possible | [25,33] |

Table 1b: Main characteristics of 3D micro/submicron surface texturing techniques of biomaterials

| Technique | Principle | Material | Pattern Type | Advantages | Disadvantages | Ref. |
|---|---|---|---|---|---|---|
| Soft lithography-Replica Molding | The replication via casting and curing of the 3D topography of a patterned, solid surface in an elastomer, usually polydimethylsiloxane (PDMS) | Polymeric materials | Dependence on the master and replication polymer used | -Simple and easy procedure (no need for clean room facilities) -Multiple stamps can be created from a single master -Reusability of the masters -High resolution (~100 nm) | Stamp distortion | [30,32] |
| Soft lithography-Microcontact printing | The pattern transfer of the material of interest-usually self assembled monolayers (SAMs) of ink-from a PDMS stamp onto the substrate surface. | Metal substrates (mainly Au or SiO$_2$- coated surfaces) | Patterning of alkanethiols, proteins, silanes, polymers, etc. onto surfaces | - Simple and easy procedure -Multiple stamps can be created from a single master -Individual stamps can be used several times with minimal degradation of performance -Both flat and curved surfaces can be stamped | -Substrate contamination -Shrinking/swelling of the stamp -Ink mobility -The resolution of the printed features, depends on the concentration of the alkanethiols and the printing conditions | [30,32] |
| Electrospinning | Fibers are electrostatically spun into a nonwoven scaffold | Polymer solution | Aligned or randomly oriented fibers | Good control over pore sizes and fiber diameters | Limited mechanical properties | [25,28] |
| Chemical Vapour deposition | Axial crystal growth formation via the VLS mechanism assisted by a metal catalyst. | Noble and transition metals | Nanowires, rods and whiskers | - High quality structures -Easier to pattern larger areas than lithography-based techniques - High aspect ratios | - Specific feature geometries not possible - Contamination from the metal catalyst | [27] |
| Etching | The creation of topographical features on a surface by selective removal of material through physical (dry etching) or chemical (wet etching) means. | Silicon, glass, plastics | Dry anisotropic etching results in a flat profile, while wet anisotropic etching results in cavities with inclined sidewalls. Isotropic etching occurs not only in the direction of depth, but also laterally, and results in a curved profile. | Wet: Clean and simple Dry: Anisotropic | The desired feature dimension depends on the etch rates Wet: Photoresist delamination Dry: Toxic gases | [16,28] |

Abbreviations: 3D: three dimensional; CAD: Computer-aided design; LIGA: German acronym for Lithographie, Galvanoformung, Abformung; VLS: Vapor-Liquid-Solid

**1.4. Different topographies used as cell culture platforms for nerve cell studies**

In an attempt to establish a link between the topographical cues with specific cell responses, the patterned surfaces must be well defined and analyzed in terms of geometry. In the case of the random surface patterns, the surface topography cannot be well defined and is usually described in terms of statistical roughness parameters, such as the arithmetic average, $R_a$ or the root mean square, $R_q$ roughness values [35]. For the geometrical description of the nanoporous surfaces, as described in 1.3, parameters such as void coverage or average pore diameter have been used.

However, the majority of the studies describe models of deterministic topographies. Such topographies will be presented in this review based on a two-stage classification. Primary classification categorizes the various pattern geometries into continuous and discontinuous. Figure 1A illustrates examples of the cross-sections of the respective topographies with the third dimension being omitted for clarity reasons. Continuous geometries can be further classified into anisotropic and isotropic topographies. Anisotropic topographies are directionally dependent, providing cues preferentially along a single axis. Examples of continuous anisotropic topographies are photolithographically fabricated grooved silicon substrates or electrospun polymer fibers at parallel orientation. Isotropic topographies provide cues along multiple axes, for example randomly-oriented fibers. In the continuous topographies the experimental parameters include feature width and depth and fiber parameters (e.g. orientation, diameter), respectively.

Discontinuous geometries are described here based on the anisotropy or asymmetry of the feature. Based on this, pillars of circular (e.g. silicon or gold pillars or post [36]) or squared cross-section are examples of isotropic or symmetric features, while cones of elliptical cross-section is an example of asymmetric or anisotropic ones [37]. In these topographies the experimental parameters include the interfeature spacing or pitch, feature height, etc. Furthermore, discontinuous topographies can further be classified based on the distribution of the features and described as periodic or random.

**Figure 1:** Classification of the deterministic topographies used as cell culture platforms for nerve cell studies. (A) Schematic illustration of the cross-sections of the various topographies, including the continuous, such as grooves and fibers (Fig. 2-4 and 7), and the discontinuous ones, such as pillars and cones (Fig. 5 and 8). The cross-section can exhibit either the same (e.g. as in the case of pillars) or differential (e.g. as in the case of cones) surface area along the feature height. (B) Classification of the deterministic topographies based on the uniformity of the features; example of continuous (top row) and discontinuous (bottom row) topographies, like grooves and pillars, respectively.

Both continuous and discontinuous topographies can be classified into uniform and graded, depending on whether they provide cues through uniform and gradual changes in the main physical feature (e.g. groove spacing) along a particular direction, respectively (Figure 1B).

However, apart from the type of the topographical feature, it has to be emphasized that the feature dimension will also play a critical role on the cell outgrowth [38]. Accordingly, feature dimensions must be correlated with the cell size or the size of the cellular extensions (such as neurites). This can provide a broader insight into the mechanisms of cell shape alteration and cellular growth in response to topography. For example in the case of grooved topographies, a groove size much larger than the cell size may promote cellular growth within the grooves. On the contrary, the same type of cells grown on top of narrow grooves could exhibit an entirely different response.

## 1.5 Interfacing nerve cells with biomaterials: The effect of surface topography

Control of the outgrowth and patterning of nerve cells and neuroglial cells via surface topography is very important in a wide spectrum of neuroscience subfields, ranging from basic research to clinical applications.

First, development of cell culture platforms with topographical cues can aid in simulating the cellular environment within a tissue in a simplified manner. *In vitro* studies with these platforms will help to address questions on how the nerve cells and the neuroglial cells respond to the cues of their environment, beyond the soluble biochemical ones. In other words, how do these cells respond either to the contact-mediated or haptotactic cues presented by the environment (e.g. ECM molecules) or to the physical characteristics of the environment *per se* (e.g. discontinuities, contours, structural patterns, morphological changes, etc.)? Topographical guidance of axons and neuroglial cells has been shown to be important during development and regeneration. Axons are guided by the morphological changes of the surrounding cells or extracellular space. For instance, during development axons can travel along pre-existing axon tracts (or fascicles) [22]. In peripheral nerve regeneration, following lesion-induced Wallerian degeneration, Schwann cells proliferate, migrate and organize themselves into longitudinal cell columns, the so called ''bands of Bungner'', which serve as a permissive substrate for the growth of regenerating peripheral nerves [39,40]. Accordingly, oriented neuroglial cells and their ECM provide indispensable pathways for guided neuronal growth [41]. Another example implicates the morphological plasticity of the astrocytes which have been reported to alter the neuronal environment and influence the astrocyte-neurons relationships [42]. Thus, studies focusing on the effect of topographical cues can provide important information on the complex processes of the axonal outgrowth and guidance and nerve-neuroglial interactions during development and regeneration.

Second, topographical guidance cues can be implemented/exploited in introducing new conceptual approaches to address old questions in neurobiology. Such an example is the design of microfluidic platforms for the study of separate nerve cell compartments and their biochemical function *in vitro*. Experiments with "Campenot chambers" which isolate nerve cell body from the periphery have provided important information on the localization of the trophic effects of the ECM molecules to induce axon elongation [43–45]. Another example of cell patterning on interfaces with topography-induced guidance cues includes the design of (multi)electrode arrays platforms for the study of the dynamics of functional neuronal networks [46]. Topographically-guided patterning of neurons on a substrate with integrated electrical contacts for each neurons can ensure a better immobilisation compared to the chemically-guided patterning [47].

Last but not least, the body of knowledge obtained from the *in vitro* studies can then be translated into the clinical neurosciences, where implantable scaffolds with incorporated topographical guidance cues can be designed in order to promote -in synergy with the biochemical cues- the *in vivo* tissue regeneration. Parallel aligned but not randomly oriented polymeric fibers -stacked into 3D configuration to into polymeric constructs- have been shown to promote axonal regeneration in a model of peripheral nerve injury, suggesting that topographical cues can influence endogenous nerve repair mechanisms [48]. Moreover, topographical guidance cues, e.g. in the form of aligned nanofibers, could be used to guide neural processes in peripheral nerve electrodes [49].

The present review article aims at summarizing the existing body of literature on the use of artificial micro- and nanoscale surface topographical features to control neuronal and neuroglial cell morphology, outgrowth and neural network topology. A number of questions arise and will be addressed:

- Which are the geometrical parameters and dimensions of the topographies that influence the cellular responses ?
- How are neuronal cells influenced by artificial topographical cues?
- How are neuroglial cells influenced by artificial topographical cues?
- Which are the mechanisms for neuronal contact guidance by topographical cues?

In an attempt to address these issues we present the different studies based on their topographical model according to the classification of section 1.4, i.e. into continuous, discontinuous and random topographies. The effect of the various topographical cues –from phenomenology to investigation of the underlying mechanisms - on the nerve cells and the neuroglial cells is presented in the chapter 2. and the chapter 3., respectively. Each chapter summarizes the studies in respective tables. The cell types, assays and materials that have been used in the presented studies are introduced in the next section.

Although the passive mechanical properties, such as the stiffness of the substrate are also very important in cell-biomaterials interactions, these are beyond the scope of the present review. There are already remarkable reviews on the effect of mechanical cues on cell processes, in general, and on neuronal development and functioning [3,50] and the reader is advised to these.

## 1.6 Cell types, assays and materials

*Cell types*

The effect of topography on neural phenotype cells has been studied on different cell types, ranging from the PC12 cell line of neuronal phenotype to the terminally differentiated primary nerve cells. PC12 cell line, which is a clonal derivative from a rat pheochromocytome, represents a neuronal cell model of terminally differentiating into neuron-like cells upon stimulation with nerve growth factor (NGF) [51]. When stimulated with NGF, PC12 cells recapitulate several steps of neuronal differentiation as they block proliferation, obtain the phenotype of sympathetic neurons -i.e. they begin to develop processes which are fine in caliber, varicose and fascicles-and become electrically excitable. Primary neurons isolated from both CNS and PNS, depending on the neuronal phenotype under study, have been cultured on different substrates. Brain neurons have been isolated either from hippocampus or the cortex in general, while sources for peripheral neurons can be dorsal root ganglia (DRG; for sensory neurons) and superior cervical ganglia (SCG; for sympathetic neurons). In particular, embryonic hippocampal cells have been extensively studied *in vitro* due to their ability to spontaneously establish a single axon from equally growing neurites enabling the study of axon initiation mechanisms with respect to topographical cues [36,52–56]. It is the most widely used system for examination of neuronal polarity [57]. Depending on the type of neurons or the developmental age, neurons need to be treated with proper growth factors that are critical for their survival (e.g. NGF for sensory neurons). Usually culture and growth of neurons on different substrates requires a specific protein coating of the surface under study, including collagen, laminin and Poly-D-lysine (PDL).

The effect of topography on cells of neuroglial phenotype has been studied mainly on dissociated primary Schwann cells [58–62], but there are few studies using Schwann cell lines, like RT4-D6P2T [63] and CRL-2765 [64]. Studies on the cellular response of astrocytes on micropatterned substrates remain limited [65,66]. Also, there is only one study reporting the culture of oligodendrocyes onto micropatterned substrates [67].

Furthermore, complex models of coculture of neurons with neuroglial cells [58,59] and organotypic models of whole ganglion explants [48,58,59,68–71] on different topographical models have been reported. Such culture systems provide a more reliable representation of the *in vivo* environment than do single cell culture ones and have been used to assess the spatial interrelations of the different cell population with respect to the underlying topography [58,59]. Finally, some studies use topographical models in order to investigate the effect of the topographical cues on astrocytes-neurons interactions in an attempt to control astrogliosis. For that reason, models of co-culture [35] or cortical cells containing both populations (e.g. hippocampal neurons and astrocytes) [34] have been used.

Assays

A great plethora of assays, ranging from cell number and orientation assays to more complex functional and differentiation assays, can be implemented to target various biological questions regarding the effect of topography on nerve cellular functions, as have been very thoroughly reviewed by *D. Hoffman-Kim et al.* [11]. Assays include immunofluorescence targeting of specific proteins, e.g. neurofilament, Tuj1, etc., or quantitatively evaluating the expression of them by analytical techniques, like western blotting. Morphological analysis of the cell shape and outgrowth but also of the exact position of the cells with respect to the patterned substrates can be assessed via Scanning Electron Microscopy (SEM). This will be very useful in order to establish a link between the topographical cues and the specific cell responses.

Differentiation and neurite outgrowth of dissociated cells with respect to the specific topographies can be evaluated via the morphological changes of the cellular shape and the polarity of the cells, i.e. multiple-branched, bipolar or unipolar. Quantitative assays include evaluation of the number of cells forming neurites [72], the neurite density (i.e. the number of neurites per cell) [73–75], the percent of cells forming neurites [37], the length of the longest neurite [54,76] and the total neurite extension length [77]. In studies with whole ganglion explants, neurite outgrowth and Schwann cell migration have been evaluated via measuring the longest distance relative to the explant edge at which the neurite has been outgrown and the Schwann cells has migrated, respectively [59,70,78].

In order to evaluate the various maturation and activation states of the Schwann cells, the expression of specific markers have been monitored, including myelin-specific genes, like MAG, P0, MBP which are significantly upregulated during Schwann cell myelination and NCAM-1 which is associated with the immature Schwann cells [62].

Materials

Different materials ranging from polymeric to metallic have been used to develop the various topographical models. Filaments or fibers have been made of synthetic polymers -like polypropylene (PP) [79], polycaprolactone (PCL) [80], poly-L-lactide (PLLA) [71], poly(acrylonitrile-co-methylacrylate (PAN-MA) [48,70] - and natural polymers (e.g. gelatin [63] and chitosan [81]). Polydimethylsiloxane (PDMS) [54,82], polystyrene (PS) [83] and poly(lactic-co-glycolic acid) (PLGA) [84] have been used for grooved patterns. Silicon has been used for the development of grooved substrates [85], pillars [36] and cones [37,58]. Gold has been used mostly for the fabrication of pillars [74]. Patterned substrates with random nanoroughness have been developed using silica nanoparticles [34] or nanoporous gold [34,86].

Moreover, the guidance effect of Schwann cells on neurons has been also exploited with the development of polymeric replicas of Schwann cell monolayer in random or aligned orientations [61]. However, these approaches are beyond the scope of this review paper and the reader is suggested in the review of *Hoffman-Kim et al.* [11].

## 2 The *in vitro* effects of topography on nerve cell morphology, functions and network topology

Topographic guidance of neurite outgrowth and axonal guidance in response to topographical cues has been *in vitro* investigated with different topographical models, in an attempt to shed light on the mechanisms of axon initiation, axonal branching and neuronal network formation. These studies are reviewed in the current chapter. Nerve cell responses –including morphology, neurite outgrowth and differentiation- to various continuous, discontinuous and random topographies reported in the literature are summarized in the Tables 2, 3 and 4, respectively, where the various studies are presented with increasing topographical feature size.

### 2.1 The effect of continuous topographies on nerve cells

*Topographic guidance of neurite outgrowth and axonal guidance*

Topographic guidance of neurite outgrowth and axonal guidance has been investigated *in vitro* using different culture substrates comprising micropatterned features at the submicron-to micron scale. Two types of geometries have mainly been reported. The first includes the anisotropic continuous geometry of alternating grooves. According to these studies, groove depth and width seem to be the critical parameters for axonal guidance and oriented neurite growth. The majority of studies have used microgrooves with a depth size ranging from 0.2 to 69 μm. It was shown that as depth increased (i.e. from 400 to 800 nm), the percentage of hippocampal neurons growing parallel to the microchannels increased (Figure 2A) [54]. Cerebral embryonic neurite outgrowth on patterns of 2 μm groove and 8 μm ridge and 1μm depth were unaffected by the underlying topography. On the contrary, when the cells were grown on the 2 μm deep patterns and the same groove/ridge width, neurites followed the orientation of the groove axis [87]. Remarkably, when the groove depth was shallow (less than 1100 nm), hippocampal neurons lost their parallel orientation tendency and grew perpendicular to the grooves [88]. The study of *Rajnicek et al*. emphasized also the effect of cell origin (i.e. developmental age and species) on the morphogenetic events of contact guidance. In particular, whereas Xenopus spinal cord neurites grew parallel to grooves independent of feature dimensions, rat hippocampal neurites regulated their direction of neurite growth depending on groove dimensions and developmental age (Figure 2B) [88].

In the studies where groove width was quite large (50-350 μm), an interesting effect of groove depth was reported: axons turned at the edges of deep grooves (22-69 μm) but not at the edges of shallow grooves (2.5-11 μm) on PDL-coated PDMS micropatterned substrates. Specifically, in response to steps of depth h =22–69 μm, the vast majority of axons appeared to be guided by surface topography by turning and either remaining inside the grooves or staying on the top surface. In contrast to this response, neurons on shallow grooves (2.5 and 4.6 μm) disregarded the topographical steps and extended axons freely into and out of the grooves (Figure 2C). An intermediate response was observed for the neurites grown on substrates with groove depth of 11 μm [82]. However, this "turning response" was lost when neurons were grown on matrigel-covered substrates containing deep grooves, emphasizing the complexity of the contact guidance phenomenon; particularly when cells were challenged with competing growth options, stimulated by biochemical and topographical cues [82].

**Figure 2: Effect of groove depth on neurite outgrowth.** (A) SEM images of hippocampal cells on cultured on 2 μm-wide and 800 nm-deep (Aa) and 1 μm wide and 400 nm deep (Ab) PDMS microchannels with immobilized NGF (0.11 ng/mm$^2$), scale bar=10 μm); (Ac-d) Graph showing the angle distribution for axons growing on 400 nm-deep (Ac) and 800 nm-deep (Ad) microchannels with immobilized NGF (0.11 ng/mm$^2$); Inset: schematic illustration of the analysis of axon length and angle (Reprinted with permission from [54]). (B) Phase contrast micrographs of E16 rat hippocampal neurons (Ba and Bb) and Xenopus spinal cord neurons (Bc and Bd), grown for 24 and 4 hours, respectively. Cells were growing on flat quartz (Ba and Bc) or on grooves of 1 μm wide and 130 nm deep (Bb) or 320 nm deep (Bd); Scale bars: 50 μm (Ba-b) and 100μm (Bc-d). (Be-f) Graph representing the orientation responses of Xenopus spinal cord (Be) and rat hippocampal (Bf) neurites on grooved substrata. The direction of neurite growth is presented as the mean percentage of total neurites parallel (white bars) or perpendicular (black bars) to the groove direction ± sd. The groove width is indicated in the upper left of each graph (Reprinted with permission from [88]). (Ca-c) Cross-sectional schematic representation (top row), and corresponding a-tubulin immunofluorescence micrographs (bottom row) illustrating typical neurons cultured on 4.6 (Ca), 11 (Cb) and 69 (Cc) μm-deep PDL-coated PDMS grooves; (Cd) Graph showing the percentage of axons that cross a step as a function of step height h. The inset shows a Hoffman DIC image of an axon from a murine cortical neuron (E13, 3 DIV); (Ce) Percentage of axons that overcome an 11 μm-high step as a function of angle of approach. The graph shows that as the angle of approach, α, increases axons are more likely to choose to bend the angle β and disregard the step, bridging onto the next plane (Reprinted with permission from [82]).

Feature, i.e. groove and ridge width is also critical. Most of the studies have used microgrooved substrates with a (groove) width ranging from 0.1 to 350 μm. Based on the feature size scale, the studies can be classified into two categories: 1) subcellular (nanometer and single-micron) and 2) cellular (tens of microns) to supracellular (up to hundreds of microns) scale. Having usually the same size scale, the one of the two features (i.e. the groove or ridge) is kept constant and the effect of the other on the cell outgrowth is being investigated. In the case of subcellular groove width, oriented growth was observed on all types of patterns, although the 100 nm wide grooves seemed to be less efficient  as compared to the patterns with larger widths, suggesting that submicron topographies- although too small to restrict growth- can also orient cellular alignment [83,89]. *Ferrari et al* studied the neuronal polarity of PC12 cells after NGF stimulation on cyclic-olefin-copolymer grooves of 350 nm depth and 500 nm groove width with varying ridge width. It was shown that cells exhibited a shift from monopolar and bipolar to multipolar morphology, as the ridge width shifted from 500 nm to 2000 nm, with 1000 nm critical dimension [89]. In another study, differentiated PC12 cells exhibited an anisotropic growth on gratings of 200 nm depth and linewidth of 500 nm and 750 nm widths, although the alignment was more striking in the narrower grooves [83].

In the cases of cellular to supracellular scale, as the width decreased, the growth orientation of the neurites was promoted in a direction parallel to channel walls and the complexity of neuronal architecture was reduced. *Yao et al.* used grooves and ridges of 5 and 10 µm width (2-3 µm deep) to study neuritogenesis in terms of neurite angle with respect to the groove axis [84]. While the PC12 cell somata were grown on top of the ridges, their neurites exhibited a preferential growth in the grooves rather on the ridges, having a significantly more parallel growth on small (5 µm) than on larger groove sizes (10 µm) (Figure 3B). *Mahoney et al.* studied the effect of grooved topography to study the effect on neuritogenesis with groove widths large enough for the PC12 cells to sit inside [75]. Using substrates of 11 µm depth and ridge/groove width ranging from 20-60 µm, it was shown that several properties of neurite growth (i.e. neurite polarity, the direction of neurite growth and the length, number and angle of neurites emerging from the cell soma) depended on microchannel width. Cells in narrower channel (20–30 µm) extended one or two neurites parallel to the channel walls. Cells in wider channels (40–60 µm) exhibited differential response, with i) the cells contacting the wall exhibiting bipolar shape and aligned, while ii) the ones not in contact with a wall exhibiting a perpendicular growth and a shift to multipolar morphology (Figure 3A) [75].

In the high-throughput study of *Li et al.* a great plethora of different topographical features of 1µm height replica-molded with PDMS have been developed and further used as cell culture platforms for hippocampal neuronal cells. These culture platforms enabled the study of multiple comparisons among the different geometrical parameters of the patterns [90]. Regarding the continuous topographical features, anisotropic features in the form of linear and circular gratings with constant groove width (i.e. 2 and 5 µm, respectively) and varying ridge width (i.e. from 2 to 15µm) have been shown to promote axonal growth with a strong guidance response (Figure 3C), which was not the case for the cell outgrowth on the discontinuous topographical features (see Section 2.2) [90]. More specifically, the guided axons were following the transition edge between ridges and grooves. Furthermore, axonal length after 2DIV was 237 µm and 214 µm in the case of the linear-based and circular-based topographies, respectively (i.e. almost 60% and 48% longer, respectively than the neurons growing on flat substrate). Another parameter that has been evaluated was branching, which was limited in the neurons grown on the gratings. A combination of increased axonal length and reduced branching supports the strong guidance effect and the directional growth of the neurons on the gratings, compared to the other topographical patterns. Moreover, discriminating between dendrites (stained for MAP2) and axons (for Tau1), it was shown that, unlike axons, dendritic growth was found to be insensitive on almost all topographical features tested in this study [90].

Though the vast majority of studies using grooved substrates report parallel alignment of neurites, other types of guidance have been reported as well [11]. When dissociated postnatal DRG sensory neurons were grown on grooved PDMS substrates of 50 μm depth and groove/ridge width ranging between 30-100 μm/30-100 μm, respectively, the majority of DRG extended neurites parallel to the groove axis, and a subset of them extended neurites perpendicular to/bridging the groove pattern. The phenomenon termed as "neurite bridging" seems to be influenced by many parameters, including cell density, groove or ridge width (Figure 3D). Study of the dynamics of bridge formation via time-lapse microscopy revealed that bridges were formed as neurites extended from a neuron in a groove, contacted adjacent plateaus, pulled the neuron up to become suspended over the groove, and the soma translocated to the ridge [53].

**Figure 3: The effect of feature width of cellular to supracellular size scale on neurite growth:** (Aa-c) SEM images of microchannels; (Ad-f) PC12 cell neurite growth in microchannels of 20 (Ad), 40 (Ae) and 60 (Af) μm width; (Ag) Table showing the average values for neurite growth of PC12 cells cultured in microchannels of different width (Reprinted with permission from [75]). (B) Orientation of neurite growth on laminin peptide-coated PLGA films. (Ba) Neurites of PC12 cells randomly orientated on a plain PLGA film; Parallel neurite growth on 5 μm (Bb) and 10 μm (Bc) grooves (Reprinted with permission from [84]). (C) Axonal outgrowth of hippocampal neuron cells on gratings with 10 μm ridge and 2 μm groove (Ca), angles with 5 μm ridge and 5 μm groove (Cb), circles with 10 μm ridge and 5 μm groove (Cc) after 2DIV, scale bar: 100μm; (Cd) Average axonal length of neurons growing on gratings, circles, dots and flat surface; (Ce) Dendrite branching index for neurons on patterns with gratings, circles, dots, and flat surface (Reprinted with permission from [90]). (Da) SEM image of DRG after 24 h culture on laminin-coated, micropatterned PDMS substrate with cell density of 125,000 cells/cm$^2$, groove/ridge width of 50/70 μm, respectively; (Db-c) Number of neurite bridges as a function of groove width (Db) and ridge width (Dc) (Reprinted with permission from [53]).

The second model used to study the topographical guidance of neurite outgrowth and axonal guidance is that of anisotropic continuous geometry of parallel oriented fibers. Research in this area is focused mainly on the effect of parallel oriented submicron-to micron electorspun fibers on neurite elongation and outgrowth (i.e. length, etc.) of single identified neuron. It was shown that sensory and motor neurites followed single fibers of submicron diameter (0.5 μm) made of PCL and PLLA, respectively and sometimes crossed the space between adjacent fibers [59,91]. Studies with PC12 cells enable the study of neurite guidance in terms of neurite orientation, length and branching. Anisotropic fibers of submicron diameters (e.g. in the range of 200-400 nm) have been shown to promote neurite guidance in terms of neurite length (compared to the random fibers) of PC12 differentiated cells. In the study of *Genchi et al*. PC12 differentiated cells exhibited higher neurite length on the parallel compared to the random collagen-coated pHB fibers, although cells on both cases exhibited the same number of neurites/cells (i.e. they were bipolar) [92]. Neurite outgrowth length was increased when the fibers were bound with protein, either physically adsorbed or covalently attached [59,77]. The majority of the studies with fibers, though, investigate the neurite outgrowth in response to the electrospun parallel fibers with a whole DRG explant, as presented in the next section.

*Directional growth of neurites from whole explants*

The effect of fiber orientation on neurite outgrowth guidance has been strongly emphasized by many studies using the model of whole DRG explant, which includes the outgrowth of axons and the migration of Schwann and other non-neuronal cells away from it. These studies investigate mainly the effect of the orientation of submicron-to micron (diameter in the range of 400-600 nm) electrospun fibers made of different polymers on neurite elongation and outgrowth (i.e.length, etc.) from the ganglion. It was reported that neurite outgrowth on parallel aligned fibers preferentially extended along the direction of the fibers, while it was randomly distributed on the randomly oriented fibers. In order to investigate the neurite alignment as a function of fiber alignment, *Corey et al.* have used a Fourier method (Fast Fourier Transform-FFT) to quantify the alignment of the PLLA fibers of ~500 ±300 nm diameter with three different orientations and the alignment of the outgrowing neurites. It was shown that neurite alignment was superior on the highly aligned fibers compared to that on the intermediate and random alignment, suggesting that fiber alignment has an important effect on neurite alignment (Figure 4A) [69]. This effect of the topography was shown to be amplified with the incorporation of biochemical factors (e.g. ECM proteins or growth factors), either immobilized on the fibers or added in the solution (Figure 4B) [93].

Although the majority of the studies use fibers of subcellular size -and more specifically in the range of submicron-to-micron size- and report neurite alignment along the direction of the fibers, the study of *Wen and Tresco* used larger fibers in order to investigate the effect of fiber diameter on neurite outgrowth. Specifically, the study investigated the neurite outgrowth using a whole DRG explant model on parallel aligned fibers of supracellular (ie. 500-1000 μm), cellular (30 μm) and subcellular size (5 μm) made of PP. They showed that neurite outgrowth (in terms of alignment, length and density) increased with decreasing diameter, suggesting that the effect of fiber diameter is very critical for neurite outgrowth and alignment (Figure 4D) [79]. Interestingly, this increase in axial alignment was accompanied by a decrease in neurite fasciculation as the filament diameter decreased. More specifically, while on 500 μm-diameter filaments, neurites appeared organized into fascicles, neurite outgrowth on the smallest values diameters (i.e. on 5 and 30 μm) was dense, spread out and uniformly distributed. This defasciculation as the filament diameter decreases was attributed to the increase of surface area giving more space for cells to adhere and grow. Precoating with ECM molecule amplified this response [79].

**Figure 4: Effect of fiber alignment on neurite orientation.** (Aa–c): SEM images of fibers with high (Aa), intermediate (Ab) and low (Ac) alignment; (Ad–f): FFT images from 256 x 256 pixel selections from the images in (Aa–c), respectively. Yellow depicts greatest intensity, blue depicts least intensity. Note that narrower areas of higher intensity (yellow-orange) in the FFT images correlate with more oriented fibers; (Ag–i): Neurites from DRG grown for 3 days on fibers of identical alignments as depicted in (Aa–c); (Aj-l): FFT images from 512 x 512 pixel selections from the images in Ag-Ai (Reprinted with permission from [69]). (Ba-b)  High-magnification confocal microscopy images of neurite morphology on random (Ba) and aligned (Bb) PLLA nanofibers;  (Bc) Quantitative measurement of neurite outgrowth on the various nanofibers (Reprinted with permission from [93]). (C) The effect of various parameters on neurite outgrowth orientation: Fluorescence micrographs showing the neurite fields of DRG cultured on aligned nanofibers under different conditions. (Ca-b) DRG cultured on scaffolds of uniaxially aligned nanofibers that were supported on bare glass coverslips. The nanofibers were collected for 1(Ca) and 15(Cb) min; (Cc-d) DRG cultured on nanofibers that were deposited on glass coverslips and then coated with different chemical species. The fiber collection time was 15 min and the scaffolds were coated with PLL (Cc)  and PLL and then laminin (Cd) from solutions with concentration 516 μg/mL; (Ce-f) DRG cultured on aligned nanofibers that were deposited on glass coverslips that have been precoated with PEG. The fiber collection time was 1(Ce) and 15 (Cf) min. The arrow in (Ca) implies the direction of alignment for the underlying nanofibers, and it applies to all other samples. All the samples were stained with antiNF200 (Reprinted with permission from [80]). (D) Effect of fiber diameter on neurite outgrowth: Confocal micrographs showing the growth pattern of DRG neurites and Schwann cells on filament bundles of varying diameter. Immunocytochemical staining for β-III tubulin, S-100 and DAPI to label cell nuclei in blue. Neurite outgrowth and Schwann cell behavior on 500 μm-diameter (Da,c) and 5 μm-diameter (Db,d,e) filament bundles (Reprinted with permission from [79]).

In contrast to the majority of the studies focusing on the effect of the fiber orientation on the neurite outgrowth orientation, the remarkable report of *Xie et al*. sheds light on the impact of other fiber parameters, i.e. fiber density, surface chemistry of the fibers and surface property of the supporting substrate. By increasing the fiber density (via the increase of electrospinning time) the neurites tended to form bundles and grow perpendicular to the direction of fiber alignment and neurite branching could be also observed (Figure 4Ca-b). However, by depositing laminin on the surfaces of the nanofibers with a relatively high density, the projection of the neurites would switch from perpendicular to parallel outgrowth (Figure 4Cc-d) [80]. In this case, the neurites were observed to adhere well on the fibers and grow along the direction of fiber alignment exhibiting increased neurite length. When DRGs were grown on fibers deposited on glass substrate, which had been pre-coated with the protein-repellant Poly(ethylene glycol) (PEG), neurons tended to grow along a direction perpendicular to the nanofibers (Figure 4Ce-f) [80]. Due to the cell repellency property of PEG, neurites tended to form fascicles to avoid the interaction with the substrate. It is the first study that reports a perpendicular neurite outgrowth with respect to the alignment of the fibers, emphasizing the complexity of axonal guidance when multiple cues are presented to the neurons.

Table 2a: The effects of artificial micro- and nanotopographies on nerve cells- Continuous Topographies

| Feature Type | Biomaterial / Fabrication Technique | Feature dimensions | Cell type / Cell assay | Treatment/ Protein Coating | Cellular response | Ref. |
|---|---|---|---|---|---|---|
| Grooves | Fused quartz / Electron beam lithography | Depth: 14-1.100 nm; Width: 1, 2 or 4 µm | Embryonic Xenopus spinal cord neurons and rat hippocampal neurons (E16&19) / Neurite growth | Hippocampal: Poly-L-lysine; Spinal neurons: None | Effect of species: Xenopus neurites grew parallel to grooves but hippocampal neurites regulated their direction of neurite growth depending on groove dimensions. Effect of feature size (depth): Hippocampal neurites grew parallel to deep, wide grooves but perpendicular to shallow, narrow ones. Effect of developmental age: The frequency of perpendicular alignment of hippocampal neurites is higher for E16 vs. E19 hippocampal neurons. | [88] |
| Grooves | Tissue culture PS / Thermal nano-imprint lithography | Depth: 200 nm; Width: 500,750 nm | PC12 cells / Differentiation after NGF | None | Effect of feature size (width): Neurite alignment was reduced with increasing width | [83] |
| Grooves | PMMA covered Si / Nanoimprint lithography | Depth: 0.3µm; Width: 0.1-0.4 µm; Spacing: 0.2-2 µm | DRGs & SCGs whole explants / Axon guidance | Matrigel | All axons grew on ridges or spacings and not in grooves. Effect of feature size (depth): On narrow grooves/ridges the axons grew on top of several ridges simultaneously, while on wider grooves/ridges the axons were found on single ridges. Effect of feature size (width): The relation of axon diameter and groove/ridge width seems to be crucial for axon guidance | [85] |
| Grooves | Cyclic olefin copolymer / Thermal nanoimprint lithography | Depth: 350 nm; Width: 500 nm; Spacing: 500-2000 nm | PC12 cells / Differentiation after NGF | | Effect of alignment: PC12 cell morphology exhibited shift from monopolar and bipolar to multipolar morphology, as the groove spacing or ridge width shifted from 5000 to 2000nm with 1000 nm critical dimension. Effective angular modulation of the FAs with groove spacing up to 750 nm. | [89] |

Table 2b: The effects of artificial micro- and nanotopographies on nerve cells- Continuous Topographies

| Feature Type | Biomaterial / Fabrication Technique | Feature dimensions | Cell type / Cell assay | Treatment/ Protein Coating | Cellular response | Ref. |
|---|---|---|---|---|---|---|
| Grooves | PDMS with immobilized NGF / Replica molding | Depth: 400, 800 nm Width: 1, 2 μm | E18 rat hippocampal cells / Polarization and axon length | | Perpendicular or parallel contact guidance in a manner dependent on surface feature sizes. Effect of feature size (depth): As depth increased, the percentage of cells growing parallel to the microchannels increased. Effect of specific binding of NGF: The topographical cues had the most pronounced effect on polarization, regardless of the simultaneous presence or absence of NGF. In contrast, axon length was increased by tethered NGF and not by topography, though an enhancing effect was seen when both cues were presented simultaneously. | [54] |
| Grooves | PLGA / Laser ablation | Depth: 2-3μm Width: 5/10 μm Spacing: 5/10μm | PC12 cells / Differentiation after NGF | Collagen type I or laminin peptide (PPFLMLLKGSTR) | Effect of feature size: More parallel growth on small than on larger groove sizes. Preferential growth on spaces and not in grooves. | [84] |
| Grooves | Photolithography | Depth: 1 or 2 μm Width: 8 μm Spacing: 20 μm | Embryonic chick hippocampal neuron (E8) / Neurite growth | Poly-L-lysine | Effect of feature size (depth): Neurites on 2 μm-groove, and 8 μm-ridged patterns of 1μm depth were unaffected. On the contrary, neurites followed the orientation of the grooves on the 2 μm- deep patterns. | [87] |
| Linear (L) & Circular (C) gratings | PDMS/ Replica molding | Width: 2 μm (L) and 5 μm (C) Spacing: 2-15 μm | Hippocampal neurons/ Neurite outgrowth | Laminin | Effect of feature type: Strong guidance effect. Axonal length was longer compared to the one on the flat substrate. Dendrite branching was limited compared to that on the flat. | [90] |
| Micro-channels | Photosensitive polyimide / Photo-lithography | Depth: 11 μm Width: 20-60μm Spacing: 10 μm | PC12 cells / Differentiation after NGF | Collagen | Effect of feature size (width): As width decreased, the growth orientation of the neurites was promoted in a direction parallel to channel walls and the complexity of neuronal architecture was reduced | [75] |

Tables 2c: The effects of artificial micro- and nanotopographies on nerve cells- Continuous Topographies

| Feature Type | Biomaterial / Fabrication Technique | Feature dimensions | Cell type / Cell assay | Treatment/ Protein Coating | Cellular response | Ref. |
|---|---|---|---|---|---|---|
| Grooves | PDMS / replica molding | Depth: 2.5 - 69 μm Width : 50-350 μm | Murine embryonic cortical neurons (E11-E14)/ Axon Guidance | Poly-D-Lysine | Effect of depth: Axons were found to cross the steps as groove depth decreased (intermediate response at 11 μm) | [82] |
| Grooves | PDMS / Replica molding | Depth: 50μm Width: 30-100 μm Spacing: 30-1000 μm | Dissociated DRGs Hippocampal neurons / Neurite bridging | | Neurite bridging formation. Effect of anisotropy: The majority of DRG extended neurites in a direction parallel to the groove pattern, and a subset of DRG extended neurites perpendicular to/ bridging the groove pattern. Effect of feature dimensions: As width increases, the number of bridges decreases as well (The highest numbers of bridges for $w_{spacing}/ w_{groove}$ =200/30) | [53] |
| Fibers | PLLA / Electrospinning | Diameter: ~524 nm | E15 rat DRG explants / Neurite outgrowth | Collagen | Effect of anisotropy: Ganglia were elongated in the direction of fiber alignment. On aligned fibers, neurites sprouted radially but turned to align to fibers upon contact, and neurite length increased on aligned fibers relative to random and intermediate fibers | [69] |
| Fibers | PCL and PCL/Collagen blend / Electrospinning | Diameter: ~0.5 μm | E10 chick DRG explants / Neurite outgrowth | No | Effect of anisotropy: The direction of neurite elongation was largely orientated in parallel with the fibers. | [59] |
| Fibers | PLLA / Electrospinning | Diameter: ~0.5 μm | P4-5 DRG explants / Neurite outgrowth | Laminin FGF immobilized | Effect of anisotropy: Neurites followed the guidance of aligned fibers and exhibited little to no branching compared to that on random fibers. Effect of biochemical cues: Significantly higher neurite outgrowth on the fibers with immobilized biochemical cues compared to the uncoated fibers | [93] |
| Fibers | PAN-MA / Electrospinning | Diameter: 400-600 nm | P1 rat DRG explants / Neurite outgrowth | No | Effect of anisotropy: The majority of neurite outgrowth from the DRGs on the aligned fiber film extended unidirectionally, parallel to the aligned fibers. | [48] |
| Fibers | PAN-MA / Electorspinning | Diameter: 0.8 μm | P1 DRG explants / Neurite outgrowth | | Effect of anisotropy: Aligned PAN-MA fibers influenced fibronectin distribution, and promoted aligned fibronectin network formation compared to | [70] |

| | |
|---|---|
| smooth PAN-MA films. Outgrown axons extended along the fibers to significantly greater extent compared to smooth films | |

**Table 2d:** The effects of artificial micro- and nanotopographies on nerve cells- **Continuous Topographies**

| Feature Type | Biomaterial / Fabrication Technique | Feature dimensions | Cell type / Cell assay | Treatment/ Protein Coating | Cellular response | Ref. |
|---|---|---|---|---|---|---|
| Fibers | Polydioxanone / Electrospinning | Diameter: ~2-3μm | E16 rat DRG explants P3 astrocytes (as glial substrate) / Neurite outgrowth | | Effect of anisotropy: Neurites grown on aligned matrices displayed directionality that mimics that of the underlying fiber orientation. DRGs cultured on a substrate of astrocytes grew more robustly and extended longer processes than when grown on a glia-free matrix | [68] |
| Filament bundles | PP / Thermal extrusion | Diameter: 5, 30, 100, 200, and 500 μm. | DRG explants / Neurite outgrowth | Fibronectin or Laminin or none | Effect of anisotropy: Neurites grew along the long axis of the filament bundles, regardless of the filament diameter. Effect of diameter: As filament diameter decreased, the neurite outgrowth was promoted in a direction parallel to the long axis of the lament bundles | [79] |
| Filaments | PLLA / Melt extrusion | Diameter: 375 μm | DRG whole explants /Neurite outgrowth & Schwann cell migration | Laminin or PolyLLysine | Effect of anisotropy: Topography-guided neurite outgrowth Effect of coating: Neurite outgrowth was significantly increased in the presence of laminin, past the Schwann cell leading edge | [71] |
| Fibers | PCL/ Electrospinning | Fiber density: 100 – 3.000 fibers/mm | E8 DRG explants / Neurite outgrowth | Laminin or coating with PEG | Effect of fiber density and coating: Both parallel and perpendicular contact guidance were observed depending on the fiber parameters | [80] |

*Abbreviations: DRG: Dorsal Root Ganglion; E18: Embryonic Day 18; NF: Neurofilament; NGF: Nerve growth factor; P1: Postnatal Day 1; PAN-MA: poly(acrylonitrile-co-methylacrylate); PCL: Polycaprolactone; PDMS: Polydimethylsiloxane; PLGA: poly(lactic-co-glycolic acid); PLLA: poly-L-lactide; PMMA: Poly(methyl methacrylate); PP: Polypropylene; PS: Polystyrene; SCG: Superior Cervical Ganglion; Si: Silicon
**Comment:** For reasons of clarity and ease of comparison, the grooved-/ridged- patterns are presented with respect to the groove (even if the authors have reported them with respect to the ridges).

## 2.2 The effect of discontinuous topographies on nerve cells

*Effect on neuron outgrowth and neurite outgrowth orientation*

In general, the effect of discontinuous isotropic topographies in the form of micro- or submicron-scale pillars, posts, cones and holes have been investigated. Generally, the feature height ranged from 1-10 μm. According to these studies, feature spacing was the critical parameter for oriented neurite outgrowth. As the spacing between pillars increased, the fidelity of alignment decreased. Hippocampal neurons were cultured on square-shaped silicon pillars of 0.5 or 2 μm diameter and spacing ranging from 0.5 to 5 μm. The neurites tended to span the smallest distance between pillars, aligning either at 0° or 90°, with the highest alignment with the larger pillars at the smallest spacing. As the spacing between pillars increased, the fidelity of alignment decreased, and at 4.5 μm spacing, the distribution of neuronal arbor was similar to that found on a flat surface (Figure 5A) [36]. In another study, hippocampal neurons have been cultured on conical posts of 10-100 μm diameter and respective height of 1/10 of the diameter and the edge-to-edge spacing ranging from 10-200 μm [55]. In this study, microcones' diameter and spacing were found to be two critical parameters in the neuron outgrowth process (Figure 5B). Neurite processes on surfaces with smaller features and smaller intercone spacings were aligned and connected in straight lines between adjacent pillars. However, as microcone diameters started to increase, neurites wrapped around the post they were already attached; this was more evident as the spacing increased beyond 40 μm. Furthermore, neurite processes on surfaces with more than 200 μm inter-feature spacing exhibited random outgrowth and wrapping. Therefore, as the feature size and spacing was increased, a transition from aligned to random growth occurred.

The importance of interpillar spacing (or pitch) for cell polarization and alignment has been strongly emphasized in the study of *Bucaro et al.* using silicon arrays of pillars with 10 μm height and varying pitch (i.e. from 0.8 to 5μm). PC12 cells exhibited polarization and alignment on all of the substrates with an interpillar distance in the range of 1.6 to 2 μm (Figure 5D) [94]. Interestingly, the same morphological cell response was to be seen on Si (Young's modulus of ∼180 GPa) and epoxy (Young's modulus of ∼1-2 GPa) substrates, suggesting that interpillar spacing may have a more significant role in inducing the specific morphological change, while the material elasticity affects only the extent of cell polarization [94]. Square arrays of silicon pillars exhibiting diameter of 461 nm, spacing of 339 nm and height of 1200 nm could support PC12 cell differentiation towards a highly branched neurite phenotype after treatment with NGF [73]. In the case of the interpillar distance at nanoscale, PC12 cell differentiation has been shown to be inhibited or limited. Specifically, gold nanopillars of 200 nm in diameter, 2 μm in height and 70 nm in spacing limited neurite outgrowth of PC12 cells exhibiting fewer and shorter neurites compared to the cells on smooth substrates [74].

*Kang et al.* used recently a model of vertically grown silicon nanowires (vg-SiNWs) of 72 ± 8 nm diameter, 7 to 10 μm length and 17.9 NW/μm² density. An accelerated polarization of embryonic hippocampal neurons at early timepoints was shown compared to the growth on the control substrates [95]. The neuronal networks formed were functional as was evaluated by somatic calcium imaging. The same group used also another topographical model of silica bead monolayer and pitch ranging from 700 to 1800 nm [96]. Increased hippocampal neurite length at early timepoints was reported with increasing pitch compared to the control substrates.

**Figure 5: Effect of discontinuous topographies on neurite outgrowth.** (Aa) Schematic representation of the different pillared topographies (with varying inter-pillar distances); (Ab) Fluorescence images demonstrate the effects of pillar width of 2.0 μm at increasing pillar gap sizes on neurite outgrowth; (Ac) SEM images of hippocampal neurons on the different substrates. SEM images show that most axons (white arrowhead) and dendrites (black arrowhead) that grew out from the cell body tended to follow the tops of pillars with widths of 2.0 μm and gaps of 1.5 μm (Ac). Processes grew at the base of pillars with widths of 0.5 μm and gaps of 1.5 μm (Ad), as well as on pillars with widths of 2.0 μm and gaps of 4.5 μm (Ae). Scale bar = 20 μm. Growth cones (black arrows) on pillars with widths of 2.0 μm and gaps of 1.5 μm exhibited a narrow profile indicative of rapid growth (Af). In contrast, growth cones (black arrows) on pillars with widths of 2.0 μm and gaps of 4.5 μm exhibited a broader profile with a few extending filopodia indicative of slower growth (Ag), Scale bar (Af-g) = 10 μm. An increase in process formation and bundling was observed on pillars with widths of 2.0 μm and gaps of 1.5 μm (Ah). Smaller processes arose from a main process at consecutive pillars, Scale bar (Ah) = 2 μm (Reprinted with permission from [36]). (Ba) Height profile of 10 μm diameter with 10 μm spacing PDMS conical post array captured using tapping mode Atomic Fluorescence Microscopy (AFM); (Bb-e) Optical micrographs of hippocampal neurons plated on glass substrates patterned with conical posts of PDMS on the pillared arrays of 10 μm diameter/10μm spacing (Bb); 20 μm diameter/ 40 μm spacing (Bc), 50 μm diameter/100 μm spacing (Bd) and 100 μm diameter/200 μm spacing (Be); scale bar: 90μm (Reprinted with permission from [55]). (C) Axonal outgrowth of hippocampal neuron cells on 10x10 μm² dots with 10 μm spacing (Ca), triangles with 5 μm space in Y direction (Cb) and flat surface (Cc) after 2DIV; scale bar: 100 μm [90]. (D) SEM images of PC12 cells grown (Da) on high-aspect-ratio Si NPs (r = 100 nm, h =10μm, p =2μm) and (Db) on an OG142 polymer micropillar array (r = 750 nm, h =10μm, p = 3.5 μm) (Reprinted with permission from [94]). (Ea-f): SEM images in tilted view (a-c) and top view (d-f) of micropatterned Si substrates of low (Ea,d), medium (Eb,e) and high (Ec,f) roughness; (Eg-i) Confocal microscopy images of neurofilament positive sympathetic neurons grown on low (g), medium (h) and high (i) roughness micropatterned Si substrates

for 6 days, scale bar: 150 μm. (Reprinted with permission from [58]).

*Li et al.* introduced a large library of continuous and discontinuous microscale size features ranging from 2 to 15 μm in size and varying pitch (from 0.5 to 20 μm) [90]. Regarding discontinuous patterns, both isotropic (i.e. in the form of rectangles and dots) and anisotropic (i.e. in the form of semi-circles and triangles) of 1 μm height and with varying feature characteristics (i.e. width, diameter) and interfeature distances were replica molded with PDMS polymer have been developed and further used as cell culture platforms to grow hippocampal neuronal cells [90]. Regarding axonal guidance, the axons were only guided along a few dots or triangles before losing tracks after a short distance (Figure 5C). After 2DIV, the axonal length on the discontinuous patterns was smaller compared to that on the continuous topographical features (183 μm - although still longer than the one on flat). Branching was lower in all the other cases of the discontinuous patterns compared to that on the flat surface [90].

We have recently explored the use of a discontinuous anisotropic surface micropattern for controlled directional outgrowth of neuronal cells and respective intracellular networks. Micropatterned surfaces had been fabricated via ultra-short pulsed laser processing on silicon and comprised arrays of microcones (MCs) that were either of elliptical or arbitrary-shaped cross-sections (Figure 5Ea-f) [58]. Although very few sympathetic neurons could grow on flat Si substrates, all micropatterned Si substrates equally supported extended neuronal outgrowth, depicting the importance of surface roughness over nerve cell outgrowth and network formation. Furthermore, neurons on the silicon MCs of elliptical cross-section and parallel orientation (medium and high roughness substrates) exhibited a preferential parallel alignment, while neurons on the MCs of arbitrary cross-section and random orientation (low roughness) formed a highly branched network exhibiting no preferred orientation (Figure 5Eg-i). Apparently, the preferential orientation of neuronal processes matched the direction of the major axis of the elliptical MCs, suggesting a dependence of the axonal outgrowth pattern on the underlying topography [58]. This topography-induced guidance effect was also observed in the more complex cell culture system of whole DRG explant, suggesting that even a discontinuous topographical pattern can promote axonal alignment, provided that it hosts anisotropic geometrical features, even though the sizes of those range at the subcellular lengthscale. The same topographical model has been used to study the PC12 differentiation after treatment with NGF. It was shown that, unlike low and medium roughness surfaces, highly rough ones exhibiting large distances between MCs did not support PC12 cell differentiation, although cells had been stimulated with NGF [37].

**2.3 The effect of random topographies on nerve cells**

Beyond the well-defined microscale topographies, there is an increasing interest on more random (or stochastic) topographies that resemble better the nanotopographies of the *in vivo* extracellular environment. *Blumenthal et al.* have used assembly of monodispersed silica colloids of increasing size in an attempt to mimic the topography from the level of receptor clusters to ECM features. Among the different substrates (i.e. corresponding surface roughness ranged from 12 to 80 nm), PC12 cells after treatment with NGF exhibited increased differentiation - in terms of polarity and neurite length- and associated functional traits on a specific $R_a$ of 32 nm (Figure 6A). Remarkably, primary hippocampal neurons also responded to roughness in a manner similar to PC12 cells exhibiting prominent, axon-like polarized structures on exactly the same roughness value of 32 nm. Furthermore, it was shown that this stochastic nanoroughness can modulate the function of hippocampal neurons and their relationship with astrocytes. Specifically, although on the rough substrates neurons were predominantly found associated with astrocytes, for a critical roughness value of $R_a$ of 32 nm, neurons were dissociated from astrocytes and continued to survive independently even up to 6 weeks (Figure 6B) [35]. Interestingly, this optimal roughness regime coincides with the roughness value of the astrocyte surface (i.e. $R_q$ = 26–28 nm) measured by AFM; this could explain the dissociation of the neurons from the astrocytes at this specific roughness regime.

In two other studies, neuronal adhesion and survival were affected in a unimodal manner when cultured on nanotextured silicon. In the first study, an intermediate value of $R_a$ = 64 nm promoted an optimal response of embryonic primary cortical neurons and both higher and lower roughness reduced this response (Figure 6C) [97]. In the study of *Fan et al*., substantia nigra neurons survived for over 5 days with normal morphology and expressed neuronal tyrosine hydroxylase when grown on surfaces with a $R_a$ ranging from 20 to 50 nm. However, cell adherence was adversely affected on surfaces with $R_a$ less than 10 nm and above 70 nm (Figure 6D) [98]. Though nanoscale materials have received a great deal of attention in recent years, it is not yet clear the role they could play in directing neuronal growth for tissue engineered applications [11].

*Cho et al.* used a model of aluminium oxide concave nanostructures with a pitch of 60 (small) and 400 (large) nm, which were fabricated by electropolishing and electrochemical anodization. The large pitch was reported to exert an accelerating effect on hippocampal neuronal polarization or axon formation [99].

**Figure 6: Effect of random topographies on neurite outgrowth.** (A) PC12 cells on nanorough substrates. (Aa)

AFM of SNP modified substrates with the corresponding surface roughness values; (Ab-d) Morphology of PC-12 cells on Rq= 3.5 nm (Ab), Rq= 32 nm (Ac), and R= 80 nm (Ad) visualized by staining for F-actin; (Ae-f) Impact of nanoroughness on PC12 polarization as assessed by determining number of neurites per cell (Ae) and neurite length (Af) (Reprinted with permission from [35]). (B) Neuron–astrocyte interaction on smooth glass (Ba) and substrate of R= 32 nm (Bb). Astrocytes and neurons were visualized using antibody against GFAP (blue) and MAP-2 (red), respectively. Quantification of neuron–astrocyte association in short-term (5 d) (Bc) and long-term (6 wk) (Bd) cultures; (Be-f) Calcium-sensitive FURA-2 imaging in hippocampal neurons on smooth glass substrates and surfaces with R of 32 nm: (Be) change in intracellular calcium level as assessed by FURA-2 intensity and (Bf) rate of depolarization as determined by the slope of the depolarization portion of the curve (Reprinted with permission from [35]). (C) Dependence of cell viability (6 days after inoculation) on $R_a$ of Si wafers [97]. (D) Dependence of cell viability (5 days after inoculation) on the $R_a$ of Si wafers (Reprinted with permission from [98]).

Table 3a: The effects of artificial micro- and nanotopographies on nerve cells- Discontinuous Topographies

| Feature Type | Biomaterial / Fabrication Technique | Feature dimensions | Cell type / Cell assay | Treatment/ Protein Coating | Cellular response | Ref. |
|---|---|---|---|---|---|---|
| Pillars | Single crystal Si / UV stepper lithography & deep reactive ion etching | Diameter: 400 nm, Length: 5 $\mu$m Spacing: 0.8 - 5 $\mu$m Aspect ratio: 12:1 | PC12 cells / Cell morphology | Pretreatment with ethanol | <u>Effect of spacing</u>: Cells exhibited polarization and alignment at 2 $\mu$m interpillar spacing | [94] |
| Square-shaped Pillars | Silicon / Photolithography | Height: 1 $\mu$m Diameter: 0.5 /2 $\mu$m Spacing: 0.5 -5 $\mu$m | Hippocampal neurons / Dendrite & Axonal outgrowth | Poly-L-lysine | <u>Effect of spacing</u>: The neurites tended to span the smallest distance between pillars, aligning either at 0° or 90°, with the highest alignment with the larger pillars at the smallest spacing. As the spacing between pillars increased, the fidelity of alignment decreased, and at 4.5 $\mu$m spacing, the distribution of neuronal arbor was similar to that found on a flat surface. | [36] |
| Conical Posts | PDMS posts on glass / Masterless soft lithography transfer method | Diameter: 10-100 $\mu$m Spacing: 10-200 $\mu$m Height: ~ 1/10(diameter) | Hippocampal neurons / Neuron process outgrowth | Poly-D-lysine | <u>Effect of feature size (diameter)</u>: On surfaces with smaller features and smaller spacings, processes aligned and connected in straight lines between adjacent pillars and mostly followed a single direction by occasionally branching in the perpendicular direction. However, as feature diameters increased, neurite wrapping around the post was observed. <u>Effect of spacing</u>: Spacing of 200 $\mu$m promoted both random outgrowth and wrapping. | [55] |
| Pillars | Gallium phosphide / CVD | Diameter: 50 nm Height: 2.5 $\mu$m | DRG neurons / Neuronal growth | | <u>Effect of topography</u>: Neuronal survival relative to control surfaces was increased, and the neurons underwent complex interactions with the nanowires, such as pulling on and bending them. | [100] |
| Dots, grids, & squares | PDMS / Replica molding | Width or Diameter: 2 - 15 $\mu$m Spacing: 0.5 - 20 $\mu$m | Hippocampal neurons/ Neurite outgrowth | Polylysine and Laminin | <u>Effect of feature type</u>: In the case of the circular and dot patterns, both axon length and branching were limited. | [90] |
| Nanowires | Au-covered Si / CVD (growth of nanowires) & Photolithography (line pattern) | Diameter: $72 \pm 8$ nm; Length: 7 to 10$\mu$m; Density: 17.9 NW/$\mu$m$^2$ | E18 rat hippocampal neurons / Neurite outgrowth & Functionality assays (calcium imaging) | Poly-L-lysine | <u>Effect of topography</u>: Accelerated polarization of the neurons compared to the growth on coverslip at early timepoints. Neurons on SiNWs formed a single, extremely elongated major neurite earlier than minor neurites. Formation of functional neuronal networks | [95] |

**Tables 3b:** The effects of artificial micro- and nanotopographies on nerve cells- Discontinuous Topographies

| Feature Type | Biomaterial / Fabrication Technique | Feature dimensions | Cell type / Cell assay | Treatment/ Protein Coating | Cellular response | Ref. |
|---|---|---|---|---|---|---|
| Pillars & Pores | Gold / Electro-deposition | Pillars diameter: 200 nm spacing: 70 nm height: 2 μm Pores diameter: 200 nm spacing: ~40 nm depth: 35 μm | PC12 cells / Differentiation after NGF | Poly-L-Lysine | Effect of feature size: PC12 cell neurite outgrowth was inhibited or limited on the pillars & pores compared to the smooth substrates | [74] |
| Conical-shaped pillars | Single crystal Si / Femtosecond pulsed laser processing | Height: ~1-2 μm | Embryonic mouse brain cells / Cell outgrowth & network formation | | Effect of surface roughness: Although only a few neurons survived on the flat substrate, the neurons on the pillars formed an elaborate web of cytoplasmic processes in the absence of glial elements. | [101] |
| Microcones (MCs) | Single crystal Si / Femtosecond pulsed laser processing | Microcones of i) arbitrary cross-section (of 2.6 μm pitch and 1.3 μm height ii) elliptical cross-section (of 4.7/6.5 μm pitch and 3/7/8.6 μm height) | Dissociated sympathetic neurons & DRG whole explants / Network formation PC12 cells / Differentiation after NGF | Collagen Collagen | Effect of surface roughness: Neurons were parallel oriented onto the MCs of elliptical shape, while they exhibited a random orientation onto the MCs of arbitrary shape. The same guidance response was to be seen to the whole DRG explants. Effect of surface roughness: Although PC12 cells had been treated with NGF, they failed to differentiate on the high-roughness substrates exhibiting large distances between MCs, independently of the MCs' chemical coatings. | [58] [37] |
| Bead monolayer | Silica | Bead diameter: 700 - 1800 nm | E18 rat Hippocampal neurons | Plasma treatment & Poly-D-Lysine | Effect of roughness/topography: Cells on the coverslip or the low diameter bead monolayer had barely sprouted neurites (at 1DIV) Effect of bead diameter: The length of the longest neurite increased with the bead diameter exhibiting a threshold at 1000nm (~70 μm at 1DIV). | [96] |

*Abbreviations: CVD: Chemical Vapor Deposition; DRG: Dorsal Root Ganglion; NGF: Nerve growth factor; PDMS: Polydimethylsiloxane; Si: Silicon.

Table 4: The effects of artificial micro- and nanotopographies on nerve cells- Random Topographies

| Feature Type | Biomaterial / Fabrication Technique | Feature dimensions | Cell type / Cell assay | Treatment/ Protein Coating | Cellular response | Ref. |
|---|---|---|---|---|---|---|
| Nano-roughness | Silica (SiO₂) nanoparticles/ Stöber process | Monodispersed silica colloids of increasing size $R_q$: 12-80 nm | PC12 cells / Differentiation after NGF Primary hippocampal neurons / Neurite outgrowth & neuron-astrocytes relationships | Collagen | Effect of nanoroughness: PC12 cells exhibited an increased differentiation- in terms of polarity and neurite length- and associated functional traits on the specific $R_q$: 32 nm. Effect of nanoroughness: Neurons exhibited prominent, axon-like polarized structures on the same $R_q$: 32 nm. Although on the rough substrates neurons were predominantly found associated with astrocytes, for this roughness value of 32 nm neurons were dissociated from astrocytes and continued to survive independently even up to 6 weeks | [35] |
| Nano-roughness | Single crystal Si / Chemical etching | $R_q$: 18, 64 & 204 nm | Embryonic rat cortical neurons/ Cell adherence | None | Effect of nanoroughness: Unimodal cell response, i.e an intermediate value of Ra = 64 nm promoted an optimal response, and both higher and lower roughness reduced this response | [97] |
| Nano-roughness | Single crystal Si / Chemical etching | $R_a$: 2-810 nm | Substantia Nigra neurons / Cell adherence and viability | None | Effect of nanoroughness: Unimodal cell response (an intermediate value range of Ra = 20-50 nm promoted an optimal response, and both higher and lower roughness reduced this response) | [98] |
| Concave & Nanoporous | Aluminium Oxide/ Electropolishing & electrochemical anodization | <u>Small concave</u> Pitch: 60 nm <u>Large concave</u> Pitch: 450 nm Nanoporous substrate with cylindrical pore channels at the center of large concave features | E18 rat hippocampal neurons / Neurite outgrowth | Coating with N-(2-aminoethyl)-3-aminopropyltri methoxysilane | Effect of pitch: Neurite development was found to be much faster on surfaces with a 400 nm pitch than on surfaces with a 60 nm pitch. Major neurites stretched out earlier from the soma and grew more vigorously on the large-concave and nanoporous substrates | [99] |

*<u>Abbreviations</u>: NGF: Nerve growth factor; $R_a$: arithmetic average roughness parameter; $R_q$: Root mean square roughness parameter; Si: Silicon

**2.4 The effect of topography on nerve cells - An insight into the mechanisms**

According to the previous sections, a considerable amount of valuable information on the nerve cell responses has been gathered from a plethora of the different topographical models (i.e. in terms of geometries, materials, cell types and assays). However, an insight into the mechanism of topography sensing of the neuronal cells is still lacking. Cell responses to topographical cues can be mainly classified into two main classes: a) The direct effect on cytoskeleton and b) the indirect effect on signaling.

In the case of non-neuronal cells it has generally been accepted that the cellular machinery that recognizes the physical and topographical characteristics of the extracellular milieu are the intrgrin-based adhesion molecules [102]. Integrin activation leads in turn to the assembly of focal adhesions (FAs) which link the integrins to the actin filaments of the cytoskeleton [103,104]. FAs include the scaffolding proteins, like vinculin or talin, which transduce forces through the actin-myosin cytoskeletal network due to cytoskeletal tension; thus they can directly result in changes in cytoskeletal organization, structure and finally cell shape [102,105]. Furthermore, FAs include the signaling proteins, such as paxillin, that initiate biochemical signal cascades [104,105]. Such cascades include the activation of phosphorylation- and G-protein (like Rho)-mediated pathways (e.g. ROCK), which can lead to long-term changes in transcriptional regulation, cell proliferation, differentiation and survival [102]. According to this hypothesis, topography can play a critical role in the way the proteins (e.g. from the serum) deposit onto the surface which then influences the cell binding via the integrin activation [104]. Especially in the case of nanotopographical features, varying features at the scale of individual adhesions may alter the clustering and reorganization of integrins thus influencing the number and distribution of FAs [106].

In the case of neuronal cells, integrins are widely expressed in the nervous system and their presence has been linked with various developmental processes such as neuronal migration, axonal growth, and guidance by attaching to ECM proteins such as laminin, fibronectin, and tenascin [107–109]. Furthermore, many of these focal adhesion proteins, such as vinculin and talin have been detected in growth cones [109,110], suggesting a role for integrin-cytoskeletal coupling in growth cone motility and neuronal processes in general. Accordingly, the majority of the studies investigating the effect of topography on nerve cells have focused on integrin activation, FAs, cytoskseleton and ROCK signaling [80,88,96].

Beyond its effect on integrin activation, topography may influence cell functions via other mechanisms, like triggering mechanosensitive ion channels or through secondary effects, such as alterations in the effective stiffness of the substrate. However, these remain still largely unexplored.

As the cellular mechanisms of topography sensing have been very explicitly discussed in the review of *Hoffman-Kim et al*. [11], this section of the present review article is focused on the *in vitro* studies with nerve cells addressing specific questions of nerve cell contact guidance and neurite outgrowth, i.e. neuronal polarity, perpendicular *vs*. parallel contact guidance and neuronal function via mechanosensitive channels, in an attempt to decipher the mechanisms controlling these responses.

*Study of neuronal polarity selection*

*Ferrari et al*. have monitored PC12 cells after NGF stimulation by time-lapse, high-resolution microscopy to evaluate neurite outgrowth and polarity during the early stages of neurite outgrowth on nanogratings. Although inititially randomly oriented neurites emerged from an unpolarized cell body, these were in time retracted, as the cell body was acquiring a spindle-like shape and aligned its two poles to the nanograting. These results suggest that the transition from multipolar to bipolar shapes was highly guided by the nanogratings. Neurites emerging from the poles explored a restricted range of alignment angles. This bipolarity and neurite alignment were maintained after reaching this polarity. Focusing on the dynamics of focal adhesion (FA) formation and maturation they showed that FAs were preferentially grown on the ridges and developed preferentially along the direction of neurite extension (i.e. FA length). This topographical constraint imposed to the FAs by the ridge width resulted in a difference in size between FAs formed at the tip of aligned neurites and those generated at the tip of misaligned ones. Accordingly, the maturation of aligned FAs was enhanced, while the maturation of the misaligned ones was lost. In an attempt to shed light on the mechanism, it was shown that this topography-driven polarity resulting into the maturation of the FAs, requires ROCK and myosin-II -mediated cell contractility. Inhibition of ROCK or of myosin-II activity affected PC12 polarity selection on nanogratings by significantly reducing the fraction of bipolar cells and hampering FA maturation [89].

*Micholt et al*. have studied the very early stages of neuronal polarity of hippocampal cells on micron-sized pillars of 3 μm height and varying width and spacing (ie. ranging from 1 to 5.6 μm and from 0.6 to 15 μm, respectively) framed by flat regions. Using time lapse imaging it was shown that the first sprout, but also the Golgi-centrosome complex, which is used as an indicator of neuronal polarization, were preferentially located on the pillar surfaces and not on the flat region [111]. This preferential orientation towards the pillars was also observed in N-cadherin distribution, whose asymmetric presence has been implicated in cell polarization [112]. All these events were independent of the inter-pillar spacing. However, growth cone morphology was shown to be dependent on inter-pillar spacing and seems to be correlated with the faster outgrowth response - i.e. longer neurites exhibited smaller growth cones- suggesting that the events of polarization and growth may not be influenced in the same way. In order to elucidate the possible signal transduction mechanism, analysis of the phosphotyrosine patches, which indicate areas of high activity for growth signaling in relation to the pillar-axons contact points, have been analyzed. Colocalization of actin filaments and phosphorylated tyrosine positions on the pillars-axon contacts suggests that tyrosine phosphorylation is involved in the regulation of actin remodeling. This behavior was dependent on the pillar spacing and was more striking on pillars with spacing in the range of 1–2 μm, which was the dimension range of the longest neurite.

*Study of perpendicular contact guidance with respect to the cytoskeleton*

*Rajnicek et al*. have attempted to elucidate the mechanism for growth cone contact guidance by physical substratum contours *via* studying the cytoskeleton elements that are implicated in the organization of the cytoskeleton on topographical cues [56]. Use of cytoskeleton inhibitors targeting the particular growth cone elements, i.e. taxol and nocodazole for the microtubules and cytochalasin B for the actin microfilaments, did not influence the parallel and perpendicular contact guidance. Furthermore, drugs that have been shown to block perpendicular neurite guidance did not affect parallel orientation, suggesting that hippocampal neurites appear to use different mechanisms for perpendicular and parallel contact guidance. In an attempt to shed light on the signaling pathway for perpendicular alignment of hippocampal neurites, inhibitors for calcium channels, for G proteins and for protein tyrosine kinases have been used. Among them, influx of calcium and protein kinase C activity were shown to be essential components in the intracellular signal transduction pathway for perpendicular contact guidance [56].

In order to shed light on the perpendicular contact guidance of DRG axons on parallel aligned fibers, *Xie. et al*. have investigated the pharmacological effect of myosin II inhibition [80]. By adding blebbistatin (i.e. an inhibitior of myosin II) to the culture medium, the perpendicular growth during the initial period of neurite extension from the ganglion would change into parallel growth. On the contrary, the presence of blebbistatin did not influence the direction of the parallel axonal growth although it had an impact on the quality of the interactions among the neurites. These results showed that the perpendicular and not the parallel neurite growth was dependent on the myosin II, suggesting an important role of myosin II in the perpendicular contact guidance of neurite outgrowth on parallel aligned fibers.

*Cytoskeletal actin dynamics are involved in pitch-dependent neurite outgrowth*

In an attempt to investigate the pitch-dependent neurite outgrowth of hippocampal neurons, *Kang et al.* have used three different biochemical inhibitors of neuronal cytoskeletal dynamics, i.e. cytochalasin D (inhibitor of the formation/function of F-actin), nocodazole (inhibitor of microtubule formation) and blebbistatin. In the case of nocodazole the neurons barely developed neurites in any substrate [96]. In the cases of cytochalasin D and blebbistatin, hippocampal neurons developed neurites; however the pitch-dependent outgrowth was lost, suggesting the importance of actin dynamics in this outgrowth response. However, treatment with the Y-27632, which is a selective inhibitor of ROCK of the Rho/ROCK pathway for cytoskeletal dynamics, did not affect this pitch-dependent neurite outgrowth. The authors suggest that biophysical (i.e. topographical) cues may induce intracellular pathways other than the Rho/ROCK pathway to influence neuritogenesis and neurite outgrowth.

*Nanoroughness modulates neuron function via mechanosensing channels*

It has been suggested that topography may influence neuron contact guidance and neurite outgrowth via other mechanisms, like triggering mechanosensitive ion channels. The study of *Blumenthal et al.* suggests a prominent role for the mechanosensitive ion channel the Piezo-1, which is expressed by CNS neurons and not by sensory neurons such as dorsal root ganglia [113], in sensing nanotopography [35]. For that reason, the expression of FAM38A, which is an integrin-activated transmembrane protein that is part of the mechanosensitive ion channel Piezo-1 has been evaluated. Whereas FAM38A expression in PC12 cells on glass was predominantly localized at neurite branch points, which would be a region of high cytoskeletal tension, FAM38A expression was distributed more homogenously in cells grown on the substrate of the nanoroughness value which supported an increased differentiation, in terms of polarity and neurite length (i.e. $R_a$=32 nm). Interestingly, DRGs, which do not express Piezo-1 channels, did not show any morphological changes on nanoroughness substrates. Furthermore, inhibition of mechanosensing cation channels including Piezo-1 diminished the increased sensitivity to depolarization that was observed in hippocampal neurons on mean roughness of 32 nm and generally influenced hippocampal neuron-astrocyte interactions on all substrates.

## 3. The *in vitro* effects of topography on neuroglial cell morphology and functions

Directional growth of neuroglial cells in response to topographical cues has been *in vitro* investigated with different topographical models, in an attempt to develop a stimulating and guiding environment for neurons. These studies are reviewed in this chapter. The current status in the literature regarding cell response –including morphology, migration and activation- to various continuous and discontinuous topographies and random roughness is summarized in the Tables 5,6 and 7, respectively, where the various studies are presented with increasing topographical feature size.

### 3.1 The effect of continuous topographies on neuroglial cells

Two types of continuous topographies in the form of i) alternating grooves/ridges and ii) aligned fibers have been mainly used to study the effect of topographical anisotropy on neuroglial cell growth and orientation.

*Growth and orientation of dissociated neuroglial cells on microgrooved substrates*

In the case of alternating microgrooves, groove width seems to be the key parameter for alignment of Schwann cells. The width of Schwann cells varies from 5 to 10 μm and pattern widths or spacings ranging from 2 to 30 μm were found to be optimal for the alignment of Schwann cells. *Lietz et al*. used PDL/LN-coated silicon chips with microgrooves of 15 μm depth and various widths (2-100 μm) and studied the effect of groove width on Schwann cell morphology. It was shown that, when cultured on the micropatterned surfaces, Schwann cells displayed polar morphologies in parallel to the microgrooves. Specifically, among the different patterned substrates tested, cells were completely aligned on microgrooves of 2-20 μm width, and this high degree of orientation was preserved over culture time (Figure 7A) [114]. In contrast, cells grown on non-structured areas appeared disorganized without preferential orientation. *Hsu et al*. used a topographical model comprising silicon grooves with depth of subcellular size (i.e. of 1.5 μm) groove width ranging from 10-20 μm. When the width/spacing of the grooves increased from 10/10 to 20/20 μm, the extent of cell alignment at 24 h was enhanced by 1.5 times (Figure 7B). Coating with laminin or collagen increased the percentage of aligned cells [115]. In another study, laminin-coated lines of varying widths or with constant width and increasing spacing (i.e. ranging from 10 to 50 μm) were patterned onto PMMA. It was shown that Schwann cells attached and became elongated along the laminin stripes. Pattern widths below 40 μm gave rise to increased degree of orientation [60].

**Figure 7: Effect of groove width on Schwann cell growth.** (Aa-c) SEM micrographs of the microgrooves of varying width (Aa); On microlanes, a longitudinal Schwann cell alignment was evident (arrowhead in (Ab)) but not on the planar surface of the same MStC (arrowhead in Ac); (Ad) Quantification of Schwann cell orientation (Reprinted with permission from [114]). (B) Optical micrographs of Schwann cells stained with anti S-100 on patterned silicon substrates at 24 h: 10/10/1.5 (Ba), 10 /20/1.5 (Bb), 20/10/1.5 (Bc) and 20/20/1.5 μm (Bd) silicon/Al (Reprinted with permission from [115]).

Apart from the parallel alignment of the cells along the groove axis, *Goldner et al.* have reported cellular bridging between neighboring microgrooves. In their study, a number of glial cell types including Schwann cells, have been shown to span grooves of a dimension larger than the cell body, from 30 to 200 μm, anchored to the plateaus only by tension bearing cellular extensions, with no underlying support in a climbing phenomenon termed "cellular bridging" [53]. These findings are very important because they suggest that cellular response to topography is more complex than simply cytoskeletal restriction.

The impact of groove depth is emphasized in the study of *Hsu et al*. using a topographical model comprising PLGA grooves with groove width and depth ranging from 10-20 μm and 0.5-3 μm, respectively. They showed that 1.5 μm was the minimal groove depth to exert a proper guiding effect on dissociated Schwann cells [115]. Coating with laminin improved short-term cell adhesion and alignment, although this effect was decaying with increasing groove depth.

Since the use of aligned neuroglial cells as guidance structures for the growing or regenerating neurons is highly envisaged, some studies investigate the guidance effect of preseeded Schwann cells on neurons. *Miller et al.* used microgrooved polymer substrates made of biodegradable PDLA of 10 μm width. Groove width was found to be a significant factor in promoting Schwann cell alignment, and widths and spacings ranging from 10-20 μm were found to be optimal [116]. The presence of preseeded aligned Schwann cells in the grooves were found to promote neurite alignment.

The topography-induced orientation effect studied with Schwann cells has also been observed in astrocytes. In particular, rat type-1 astrocytes have been shown to get aligned along the direction of the groove of 3 and 10 μm depth and width, respectively. Cell adhesion and spreading of cytoskeletal filaments were significantly improved on the laminin coated substrates at all seeding densities tested [66]. Astrocytes have been shown to orient themselves even in shallower grooves of 250 nm depth (width: 1 μm) [117].

Oligodendrocytes were also sensitive to shallow grooves. In the remarkable study of *Webb et al.* the effect of anisotropic topographical cues on oligodendrocyte progenitor and mature oligodendrocytes has been studied and compared with that on hippocampal neurons. For that reason, quartz grooves of varying depth (i.e. 0.1-1.17 μm), width (i.e. 0.13-4 μm) and spacing (i.e. 0.13-8 μm) have been used. Both cell types of oligodendroglial lineage were found to be highly aligned by all topographical contours, although this alignment was less striking as inter-groove spacing increased to several micrometres.  Remarkably, cells aligned even in the ultrafine topographical cues down to 100 nm, which was not the case for the neurons, suggesting the high degree of sensitivity of the oligodendrocyte lineage to topographical anisotropy [67]. To our knowledge, there is no additional study on the effect of oligodendrocyes onto micropatterned substrate.

*Growth and orientation of dissociated neuroglial cells on fibrous substrates*

Studies with parallel aligned electrospun polymeric fibers emphasize the importance of anisotropy in cell orientation. When Schwann cells were cultured on aligned electrospun fibers of submicron-to-micron diameter (0.3-1 μm), both cell cytoskeleton and nuclei were elongated and aligned along the fiber axes. On the contrary, cells cultured on random fibers exhibited random orientation [62,71]. *Gnavi et al.* showed that both primary Schwann cells and cells of a Schwann cell line were elongated along the axis of the aligned gelatin fibers of ~200-250 nm. Interestingly, cell adhesion and proliferation were deteriorated on the aligned *vs.* the random fibers [63]. When Schwann cells were cultured on aligned electrospun chitosan fibers of wider diameter (i.e. 15 μm), they migrated via spiraling along the fibers and exhibited two shapes: spherical and long olivary. Furthermore, it was observed that the long olivary cells inclined to encircle chitosan fibers up in a 3D fashion after 14 days [81].

Remarkably, this Schwann cell morphology change described above seems to be accompanied by a change in maturation and functionality. Two studies report that, compared to random and control unpatterned substrates, aligned fibrous scaffolds significantly up-regulated the expression of early myelination markers (i.e. myelin-associated glycoprotein and myelin protein zero, the cell adhesion molecule neural cadherin and the extracellular matrix molecule neurocan). This was accompanied by a down-regulation of non-myelinating immature Schwann cell markers (i.e. neural cell adhesion molecule) [64].

The effect of anisotropy imposed by parallel aligned fibers on cell directionality has been also reported with astrocytes [68]. *Cao et al.* have grown rat astrocytes on randomly oriented electrospun polymeric fibers of 665 nm diameter and studied their morphology, proliferation and activation. Compared to flat surfaces (i.e. in the form of a film), astrocytes on the fibers showed suppressed proliferation and increased apoptosis, exhibiting elongated and ramified cell shapes [65]. Furthermore, the Chondroitin polymerization factor (ChPF) siRNA uptake efficiency of the cells has been investigated for their potential on gene-silencing applications, since silencing ChPF has been shown to promote neurite outgrowth. An enhanced siRNA silencing efficiency of the astrocytes grown on the nanofiber scaffolds was achieved compared to the polymer film.

*Directional Schwann cell migration from whole explants on fibrous substrates*

Electrospun polymeric fibers have been also used to study the guidance effect of anisotropy on the Schwann cell migration from whole DRG explants. In these studies, fibers of parallel aligned and/or random orientation with constant fiber diameter of submicron (i.e. up to 500 nm) to micron (i.e. up to 3 μm) scale were developed and the effect of anisotropy on Schwann cell migration was investigated. According to these studies, although cells grown on random electrospun polymer matrices showed no directional preference, when the explants were grown on aligned matrices, Schwann cell migration displayed directionality that mimics that of the underlying fiber orientation [48,59,69–71]. Additionally, Schwann cells demonstrate the bipolar phenotype seen along the fiber.

In another study, *Wen and Tresco* reported the effect of filament diameter on DRG outgrowth using PP fibers of varying diameter ranging from supracellular and beyond (500 to 100 μm), cellular (30 μm), down to subcellular size (5 μm). An increasing degree of alignment of the Schwann cells parallel to the long axis of the filament bundles was observed with decreasing filament diameter, with Schwann cells being highly aligned along the 5 μm filaments [79]. Treatment of the filaments with laminin or fibronectin increased the maximum migration distance of the cells in all types of fibers. This amplifying effect was more obvious while the fiber diameter was decreased.

**Tables 5a:** The effects of artificial micro- and nanotopographies on neuroglial cells- **Continuous Topographies**

| Feature Type | Biomaterial / Fabrication Technique | Feature dimensions | Cell type / Cell assay | Treatment/ Protein Coating | Cellular response | Ref. |
|---|---|---|---|---|---|---|
| Sinusoidal grooves | Azobenzene copolymer / holographic surface relief grating | Depth: 250 nm Width: 1 µm | Primary human astrocyes | | Effect of topography: Cells were preferentially attached onto the pattern rather than the flat region, | [117] |
| Grooves | Quartz / Laser holographic method | Depth: 0.1-1.17 µm Width: 0.13-4 µm Spacing: 0.13-8 µm | Oligodendrocyte progenitor, mature oligodendrocytes & astrocytes | Poly D Lysine | Effect of anisotropy: Cells of the oligodendroglial lineage were found to be highly aligned by all topographical contours. Effect of feature width: As inter-groove spacing increased to several micrometres, both cell types of oligodendroglial lineage became less aligned by the patterns. | [67] |
| Grooves | Silicon / Microlithography PLGA / Soft lithography | Depth: 0.5-1.5 µm Width: 10-20 µm Spacing: 10-20 µm | Rat Schwann cells | Collagen or laminin | Effect of anisotropy: Cell bodies on patterned surfaces were more elongated compared with the rounder cells observed on the unpatterned substrates Effect of width & spacing: When the width/spacing of the grooves increased from 10/10 to 20/20 µm, the extent of cell alignment was enhanced by 1.5 time Effect of groove depth: The minimal groove depth to provide a proper guiding effect was 1.5 µm Effect of protein coating: Coating with laminin improved short-term cell adhesion and alignment, although this effect declined with increasing groove depth. | [115] |
| Grooves | PDLA / Compression molding or solvent casting | Depth: 1-4 µm Width: 10 µm Spacing: 10/20 µm | Rat Schwann cells / Effect on neurite alignment | Laminin | Effect of anisotropy: Preseeded aligned Schwann cells promoted neurite alignment. | [116] |

**Tables 5b:** The effects of artificial micro- and nanotopographies on neuroglial cells- **Continuous Topographies**

| Feature Type | Biomaterial / Fabrication Technique | Feature dimensions | Cell type / Cell assay | Treatment/ Protein Coating | Cellular response | Ref. |
|---|---|---|---|---|---|---|
| Grooves | PS / Solvent casting | Depth: 3μm Width: 10 μm Spacing: 20 μm | Rat type-1 astrocytes | Laminin solution | Effect of anisotropy: While astrocytes on unpatterned substrates were randomly oriented, most of the cells on the patterned substrates were aligned in the direction of the grooves. Effect of coating: Cell adhesion and spreading of cytoskeletal filaments were significantly improved on the laminin coated substrates. | [66] |
| Grooves | Silicon / Deep reactive ion etching | Depth: 15 ±2 μm Width: 2-100 μm | Rat Schwann cells from sciatic nerves | Poly-D-lysine / Laminin | Effect of anisotropy: Cells displayed polar morphologies in parallel to lanes, whereas cells on non-structured areas appeared disorganized without preferential orientation. Effect of groove width: Cell orientation was increased with decreasing width. | [114] |
| Lines | PMMA / Microcontact printing | Width: 10-50 μm Spacing: 10-50 μm Type I graduated line widths and/or spacing Type II constant line and spacing width | Rat Schwann cells | Oxygen plasma treatment / Laminin | Effect of anisotropy: Schwann cells attached and elongated along the laminin stripes. Effect of width: Smaller pattern widths (<40 μm) increased the degree of orientation. Effect of culture time: No significant change in orientation was observed between the two time points (ie. 4 h and 4 days). | [60] |

Table 5c: The effects of artificial micro- and nanotopographies on neuroglial cells- Continuous Topographies

| Feature Type | Biomaterial / Fabrication Technique | Feature dimensions | Cell type / Cell assay | Treatment/ Protein Coating | Cellular response | Ref. |
|---|---|---|---|---|---|---|
| Grooves | PDMS / Replica molding | Depth: 30/60 µm Width: 30/60 µm Spacing: 30/60 µm (With a slight curvature) | Schwann cells from adult rat sciatic nerve / Time lapse microscopy | Plasma treatment / Poly-L-lysine | Effect of anisotropy: Schwann cell directionality and parallel direction velocity were higher on microgrooved compared to flat substrates. Effect of width: Improved alignment on the narrower features Effect of feature type: Cells on grooves traveled faster than those on plateaus. Cells on grooves spent significantly more time contacting the feature edges than cells on grooves spent contacting the feature walls. | [118] |
| Grooves | PDMS / Replica molding | Depth: 50µm Width: 30-200 µm Spacing: 30-1000 µm | Schwann Cells from rat sciatic nerves | Poly-L-Lysine | Formation of cellular extension bridging (at d/w/s: 50/60/60 µm) | [53] |
| Fibers | Gelatin (GL/PEO_GPTMS) / Electrospinning | Diameter: 238.9 ± 74 nm | 1)RT4-D6P2T Schwann cell line 2)primary adult Schwann Cells from sciatic nerves | No | Effect of anisotropy: Cells on the aligned fibers were elongated, exhibiting aligned actin filaments. Cell adhesion and proliferation was decreased on the aligned compared to the random fibers. | [63] |
| Fibers | PLGA / Electrospinning | Diameter: 350 ±50 nm | Rat Schwann cells (CRL-2765) | No | Effect of anisotropy: Cells on random fibers exhibited spherical shape whereas cells on aligned fibers showed spread morphology along the direction of alignment of the nanofibers. Cells on aligned fibers significantly up-regulated the expression of early myelination markers old compared to film and random fibers | [64] |
| Fibers | PAN-MA / Electrospinning | Diameter: 400-600 nm | P1 rat DRG explants / Schwann cell migration | No | Effect of anisotropy: The majority of Schwann cell migration from the DRGs on the aligned fiber film extended unidirectionally, parallel to the aligned fibers. On the contrary, the orientation of Schwann cell migration on the random fiber films was randomly distributed | [48] |

**Table 5d:** The effects of artificial micro- and nanotopographies on neuroglial cells- Continuous Topographies

| Feature Type | Biomaterial / Fabrication Technique | Feature dimensions | Cell type / Cell assay | Treatment / Protein Coating | Cellular response | Ref. |
|---|---|---|---|---|---|---|
| Fibers | PCL & PCL/Collagen blend / Electrospinning | Diameter: ~0.5 µm | E10 chick DRG explants / Schwann cell migration | | Effect of anisotropy: The direction of cell migration was largely orientated in parallel with the artificial fibers. | [59] |
| Fibers | PLLA / Electrospinning | Diameter: ~524 nm | E15 rat DRG explants / Schwann cell migration | | Effect of anisotropy: Schwann cells adhering to planar glass exhibited good spreading, while those adhering to fibers elongated and became extremely narrow | [69] |
| Fibers | PCLEEP / Electrospinning | Diameter: 665 ±11 nm | Rat cortical astrocytes / Cell growth, morphology and siRNA uptake efficiency | No | Effect of topography: Astrocyte proliferation and apoptosis on the fibrous substrates was decreased and enhanced, respectively compared to the film. Fiber topography enhanced astrocyte elongation and stellation. Gene-silencing efficiency was enhanced on the fibers | [65] |
| Fibers | PAN-MA / Electrospinning | Diameter: 0.8 µm | P1 DRG explants and dissociated Schwann cells / Schwann cell migration and ECM organization | | Effect of anisotropy: Aligned PAN-MA fibers influenced fibronectin distribution, and promoted aligned fibronectin network formation compared to smooth PAN-MA films. | [70] |
| Fibers | PCL / Electrospinning | Diameter 1.03 ± 0.03 µm (aligned) 2.26 ± 0.08 µm (oriented) | Primary human Schwann cells from fetus sciatic nerves/ Cell morphology and gene expression | No | Effect of anisotropy: Cells followed the orientation of the underlying fibers. When cells grew on aligned fibers, the cytoskeleton and nuclei aligned and elongated along the fiber axes. Cells on both random and aligned fibers compared to the control substrate (film) were directed towards a pro-myelinating state. Gene expression of the neurotrophic factors was down-regulated on the electrospun fibers compared to the film. | [62] |

Table 5e: The effects of artificial micro- and nanotopographies on neuroglial cells- **Continuous Topographies**

| Feature Type | Biomaterial / Fabrication Technique | Feature dimensions | Cell type / Cell assay | Treatment/ Protein Coating | Cellular response | Ref. |
|---|---|---|---|---|---|---|
| Fibers | Polydioxanone / Electrospinning | Diameter: ~2-3 μm | P3 rat astrocytes / Cell morphology | | Effect of anisotropy: Astrocytes grown on random matrices showed no directional preference, whereas astrocytes grown on aligned matrices displayed directionality. | [68] |
| Fibers | Chitosan | Diameter: 15 μm | Rat Schwann cells | No | Effect of anisotropy: Schwann cells grew onto chitosan materials with two different shapes: spherical and long olivary. | [81] |
| Filament bundles | PP / Thermal extrusion | Diameter: 5-500 μm | DRG explants / Schwann cell migration | Laminin or fibronectin | Effect of diameter: As filament diameter decreased, Schwann cell alignement parallel to the long axis of the filament bundles was more remarkable.<br>Cell migration distance decreased with increasing fiber diameter on both uncoated and protein-coated fibers.<br>Effect of protein coating: Cell migration was enhanced equally by FN and LN | [79] |
| Filaments | PLLA / Melt extrusion | Diameter: 375 μm | DRG whole explants / Schwann cell migration | Laminin and/or PLL | Effect of anisotropy: Topography-guided neurite outgrowth<br>Effect of coating: Schwann cells were found to grow on all types of filaments. | [71] |

*Abbreviations: DRG: Dorsal Root Ganglion; E18: Embryonic Day 18; LN: Laminin; FN: Fibronectin; NF: Neurofilament; NGF: Nerve growth factor; P1: Postnatal Day 1; PAN-MA: poly(acrylonitrile-co-methylacrylate); PCL: Polycaprolactone; PCLEEP: poly[caprolactone-co-(ethyl ethylene phosphate)]; PDMS: Polydimethylsiloxane; PLGA: poly(lactic-co-glycolic acid); PLLA:poly-L-lactide; PMMA: Poly(methyl methacrylate); PP: Polypropylene; PS: Polystyrene; SCG: Superior Cervical Ganglion; Si: Silicon
**Comment:** For reasons of clarity and ease of comparison, the grooved-/ridged- patterns are presented with respect to the groove (even if the authors have reported them with respect to the ridges).

## 3.2 The effect of discontinuous topographies on neuroglial cells

*The effect on cell growth and orientation*

Although the effect of anisotropic continuous geometries on Schwann cell alignment has been extensively studied, there are not many studies reporting on the engineering of cell alignment with discontinuous geometries. We have recently suggested that even a discontinuous topographical pattern can promote Schwann cell alignment, provided that it displays anisotropic features in a parallel orientation with interfeature distance at the subcellular level [58]. We have particularly shown that the parallel oriented microcones (MC) of elliptical cross section promoted oriented Schwann cell outgrowth along the major axis of the ellipse. Interestingly, Schwann cells followed and aligned with the orientation of this MC feature, which was more pronounced as the total roughness increased. On the contrary, cells on low roughness substrates, exhibiting a discontinuous but random pattern, showed an isotropic orientation, resembling to the growth pattern of the cells grown on flat surfaces (Figure 8A). This guidance effect was also observed in the migrating Schwann cells out of the DRG explants [58].

*Lee et al.* cultured an astrocyte cell line (C6 glioma cells) onto nanodotted-arrays with dot diameter ranging from 10 to 200 nm. They showed that cell viability, morphology and adhesion varied on the different substrates and exhibited optimal cell growth on 50 nm nanodots. Except for the increased cell surface area, cells on the substrate with 50 nm nanodots exhibited also a highly branched and complex network (syncytium), which is considered as a structural support for neurons with respect to cell-to-cell signaling (Figure 8B) [119]. In an attempt to provide an insight into the impact on the interecellular communication on the various substrates, it was shown that the expression level and cellular transport of the gap junction protein Cx43 in C6 glioma cells varied depending on the diameter of the dotted-arrays used.

**Figure 8: Effect of discontinuous topographies on neuroglial cell growth.** (A) Effect on Schwann cells: (Aa) Confocal microscopy images of S100b positive Schwann cells grown on different silicon substrates for 5 days of culture, scale bar: 150 μm; (Ab) Schwann cell orientation expressed in terms of the orientation angles' (frequency) distribution. The orientation angle of the nucleus, that is approximated as elliptical, was measured as the angle between the major axis of the ellipse and the vertical axis of the image plane. The number of cells exhibiting an orientation angle value within a specific range is expressed as percentage of cells ± standard error of the mean (Reprinted with permission from [58]). (B) Effect on glioma cells: (Ba) Confocal microscopy images of C6 glioma cells seeded on nadot arrays and incubated for 24, 72, and 120 h (green: vinculin and red: GFAP; Scale bar: 25μm); (Bb) Graph illustrating the density of branching (left) and the density of the meshes (right) against the diameter of the nanodots and grouped by incubation time (Reprinted with permission from [119]).

### 3.3 The effect of random topographies on neuroglial cells

*The effect on cell growth and orientation*

Topographical models of random nanoroughness have been used for the study of neuron-astrocyte interactions in order to shed light on the mechanisms of astrogliosis. *Blumenthal et al.* studied the hippocampal neurons-astrocytes interactions onto substrates of increasing nanoroughness made of monodispersed silica colloids. Although neurons on the rough substrates were predominantly found associated with astrocytes, there was a critical roughness value of $R_q$= 32 nm at which neurons were dissociated from astrocytes and continued to survive independently even up to 6 weeks [35]. In an attempt to investigate this decoupling of the hippocampal neurons from the astrocytes, the morphology of the astrocytes on the different roughness substrates has been investigated. Interestingly, astrocytes onto the critical roughness value of 32 nm exhibited a more migratory phenotype, accompanied by a significant increase in astrocyte surface roughness (Figure 9A). This remarkable finding provides a direct link between the physical properties of the cell surface and the underlying substrate roughness.

*Chapman et al.* reported the novel ability of nanoporous gold (np-Au) to reduce astrocytic coverage through topographical cues. Specifically, they used a model of npAu comprising ligaments of average width of ~30.6 to 88.6 nm and pores of diameter: ~87-149 nm. Using an *in vitro* neuron−glia co-culture model, they showed that np-Au selectively suppressed astrocytic coverage while maintaining high neuronal coverage [34]. This finding was independent of ligament width and pore diameter. By systematically discriminating between chemical and topographical effects, the cytotoxicity of the substrates as a possible culprit for the astrocytic reduction was excluded. In an attempt to give an insight into the possible mechanism, it was shown that the np-Au surface topography inhibits the initial spreading of astrocytes across the material surface compared to the unstructured planar Au (Figure 9B).

**Figure 9: Effect of random topographies on astrocytes growth.** (A) Nanoroughness alters physical attributes of astrocytes: (Aa) Graph showing the form factor of astroyctes on the rough substrates of increasing nanoroughness. Decrease in form factor on $R_q$= 32 nm surfaces is consistent with a more motile phenotype (Inset); (Ab) Morphological changes to astrocyte cell surface on nanorough surfaces: AFM images of astrocytes grown on glass (left) and on $R_q$ = 32 nm surface (right) (Reprinted with permission from [35]). (Ba) SEM images of four different np-Au surface morphologies (M1−M4) prepared by dealloying for 15 min to 24 h; (Bb) High magnification (40x) images of neurons and astrocytes at *in vitro* day 1, 7, and 12 on np-Au and unstructured pl-Au.  Astrocytes are highlighted by a red outline in the *in vitro* day 1 and 7 images. Astrocytes and neurons were visually differentiated through colocalization of tubulin-βIII and GFAP (Reprinted with permission from [34]).

Table 6: The effects of artificial micro- and nanotopographical features on neuroglial cells- **Discontinuous topographies**

| Feature Type | Biomaterial / Fabrication Technique | Feature dimensions | Cell type / Cell assay | Treatment / Protein coating | Cellular response | Ref. |
|---|---|---|---|---|---|---|
| Nanodots | Tantalum oxide / Anodic Aluminium Oxide processing | Dot diameter: 10-200 nm | C6 glioma-astrocytoma rat cell line/ Growth, networking (syncytium) | No | Effect of diameter: Cell viability, morphology, cytoskeleton, and adhesion showed optimal cell growth on 50-nm nanodots. Cell syncytium exhibited maximum complexity on the 50nm nanodots. Expression level and cellular transport of the gap junction protein Cx43 protein in C6 glioma cells varied on the different diameter dotted-arrays. | [119] |
| Microcones | Single crystal Si / Femtosecond pulsed laser processing | Microcones of i) arbitrary cross-section (of 2.6 µm pitch and 1.3 µm height ii) elliptical cross-section (of 4.7/6.5µm pitch and 3/7/8.6 µm height) | Dissociated rat Schwann cells from sciatic nerves & mouse DRG whole explant/ Cell outgrowth & Cell migration | Collagen | Effect of anisotropy: Although arbitrary cell growth takes place on the micropatterned surfaces with microcones (MC) of arbitrary cross-section, cells were parallel aligned along the major axis of the MC on the substrates with MCs of elliptical cross-section. The same topography-guidance response was observedat the whole DRG explants. | [58] |

Table 7: The effects of artificial micro- and nanotopographical features on neuroglial cells- **Random topographies**

| Feature Type | Biomaterial / Fabrication Technique | Feature dimensions | Cell type / Cell assay | Treatment / Protein coating | Cellular response | Ref. |
|---|---|---|---|---|---|---|
| Nano-roughness | Silica (SiO$_2$) nanoparticles/ Stöber process | Monodispersed silica colloids of increasing size $R_q$: 12-80nm | Primary hippocampal neurons and astrocytes / Neuron-astrocytes relationships | Collagen | Effect of nanoroughness on neurons-astrocytes interactions: Although on the rough substrates neurons were predominantly found associated with astrocytes, at $R_q=32$ nm neurons were dissociated from astrocytes and continued to survive independently even up to 6 week. Effect of nanoroughness on astrocytes: Nanoroughness altered the apical roughness of healthy astrocytes. | [35] |
| Nano-porous | Gold / Alloy corrosion process | Average ligament width (nm): 30.6 ± 1.24 to 88.61 ± 4.89 Average pore diameter (nm): 87.11 ± 4.55 to 149.24 ± 9.9 | Rat cortical cells / Neuron-glia coculture model | Some samples were coated with Al$_2$O$_3$ film | Effect of nanoroughness on neurons-astrocytes interactions: Np-Au selectively suppressed astrocytic coverage while maintaining high neuronal coverage, independent of ligament width and pore diameter. Effect of nanoroughness on astrocytes: Np-Au surface topography inhibits the initial spreading of astrocytes across the material surface compared to the unstructured planar Au. | [34] |
| Nano-porous | Gold / Dealloying process | Void coverage: 0 to 32.5±0.8 % Average void area: 0 to 23700±2000 nm$^2$ | Murine astrocyte cells | Air plasma treatment | Effect of nanoroughness: While average cell area was inversely proportional to percent void coverage, it displayed a non-monotonic dependence to average void area. | [86] |

### 3.4 The effect of topography on neuroglial cells: An insight into the mechanisms

In the above sections the morphological and functional responses of neuroglial cells on various topographical models have been presented. However, an insight into the mechanism is still lacking. Although there is an increasing number of studies investigating the mechanism mediating the contact guidance effects of cell alignment on nerve cells (Section 2.4) either by studying the cytoskeleton elements that are implicated in the organization of the cytoskeleton on topographical cues, or by studying the indirect effect on signaling, similar studies with neuroglial cells are still lacking.

In an attempt to shed light on the Schwann cell motility on microgrooved and flat substrates, *Mitchell and Hoffman-Kim* studied the effect of anisotropic topography on rat Schwann cells cultured on laminin-coated microgrooved PDMS platforms of 30 or 60 μm groove width. Timelapse microscopy monitored cell trajectories revealing information about the cell motility and preference between the groove and plateaus [118]. The interfeature distance was shown to be important, while the smaller distance 30 μm was shown to reduce the prevalence of a multipolar morphology, giving rise to a more oriented movement. Through analysis of the velocities, they found that the parallel direction velocity was higher for the the microgrooved *vs*. the flat substrates while the reverse effect was recorded for the perpendicular direction velocity (Figure 10b). However, the overall cell velocity was relatively similar regardless of the topographical cue. The authors suggest that Schwann cells may respond to topography by orienting without altering their motility machinery. Furthermore, plateaus and groove floors exhibited distinct cues promoting differential motility and variable interaction with the topographical features. More specifically, it was shown that Schwann cells on grooves traveled faster than those on plateaus, where the cells spent significantly more time contacting the feature edges than contacting the feature walls on the grooves (Figure 10c). These findings imply that Schwann cells are initially oriented by the topographical cues, but this alignment can be preserved with occasional or periodic additional contact with the guiding feature, suggesting that the establishment of polarity may be a significant feature of the contact guidance mechanism in this topographical model. The complexity in motility was accompanied by a complexity in the phenotype, according to which an individual Schwann cell could exhibit multiple distinct motile phenotypes, with a rare unipolar morphology being correlated with a significantly increased velocity (Figure 10d) [118].

**Figure 10: Effect of groove width on Schwann cell growth.** (a) Schematic illustration of top and side

views of culture platforms, consisting of alternating raised (plateau, P) and indented (groove, G) regions. Arrows indicate directionality in motility studies: movement parallel to the topography occurs along the x axis, and movement perpendicular to the topography occurs along the y axis. Cells on plateaus encounter edge topography, while cells on grooves encounter corner and wall topography; (b) Parallel (x) and perpendicular (y) components of cell velocity on flat and microgrooved substrates; (c) Graph showing Schwan cell velocity as a function of the number of extensions (0–4). For all conditions, Schwann cells exhibiting one extension (unipolar morphology) migrated with a significantly higher velocity than cells with zero, two, three, or four extensions; (d) Phase contrast micrographs show cells with unipolar, bipolar, and multipolar morphologies (Reprinted with permission from [118]).

# 4. Discussion and Conclusions

In the previous sections, the nerve and neuroglial cell responses to the three topographies (i.e. continuous, discontinuous and random topographies) have been presented. In the following, the results from these studies will be compared and correlated. It has to be emphasized that any attempt to compare the results among the various studies using similar topographies must take into consideration the specific cell subtypes used (e.g. PNS or CNS neurons/neuroglial cells; primary or cell lines), but also the age of the animal from which the cells had been isolated.

## 4.1 Effect on neurons

Continuous topographies in the form of (i) alternating grooves/ridges and (ii) parallel aligned fibers have been shown to strongly enhance axonal guidance and orientation along groove and fiber axis, respectively.

In the case of the grooved patterns the groove depth and width are shown to be the critical parameters for studying axonal guidance/oriented neurite outgrowth and neuronal polarization/branching, respectively [37,75,87–89]. For grooves of subcellular width size, the respective studies show that the majority of the hippocampal neurons and the PC12 cells can orient parallel to the groove axis provided that the topographical features are at least of 800 nm/2 μm and 200 nm /500 nm (depth/width), respectively [54,87,88]. While, for grooves of cellular to supracellular width size, where neurons can grow in the channels among the grooves, the growth orientation of the neurites is promoted towards a direction parallel to channel walls with decreased neuronal branching as the width is decreased. In shallow grooves (i.e. up to 11 μm height), neurons could cross the grooved step, while on deep (i.e. up to 50 µm height) grooves sensory neurons have been shown either to align along the groove axis or to bridge among grooves in the absence of underlying solid support [82]. This variety of distinct responses emphasizes the dynamic and complex nature of the topography-induced neuronal growth and axonal guidance.

Concerning the parallel oriented fibers, strong topography-induced axonal guidance can be obtained using fibers of submicron-sized diameter (i.e. 200-500 nm), which resembles the diameter of the neurites [59,69,77,91–93]. This topography-induced guidance effect is observed on both dissociated primary neurons and in more complex cell culture systems, such DRG explant, and is amplified with the incorporation of biochemical signals factors (e.g. ECM proteins or growth factors) covalently bound on the fiber surface [93]. Although the strong axon alignment has been correlated with the underlying fiber alignment, recent studies demonstrate that the neurite orientation can be influenced by a set of additional parameters, including the surface chemistry of the fibers or the substrate and the fiber density [80]. Furthermore, axonal outgrowth fasciculation seems to be affected by the fiber diameter/thickness. DRG neurons on polymeric fibers of cellular to supracellular dimension showed decreased alignment and increased fasciculation compared to the subcellular size fibers (i.e. 5 µm). These results emphasize that an ensemble of neuronal processes (i.e. orientation, parallel contact guidance, fasciculation) can be topography-guided and provide new guidelines in designing fiber-based scaffolds for peripheral nerve regeneration.

Besides continuous topographies, substrates with isotropic discontinuous topographies in the form of pillars and cones of subcellular scale can control the outgrowth of neuronal processes [36,55,73,90,94]. In these topographical models, neurons grow on top of the features forming extensive 2D networks and the feature spacing or pitch seems to be the critical parameter for oriented neurite outgrowth, with a spacing range of ~0.5-3 µm for optimal alignment [36,94]. Neurons on discontinuous topographical features of larger feature sizes (i.e. tens to hundreds microns) exhibit wrapping around the feature promoting the formation of networks of 3D architecture [55]. In this case, both feature size and interspacing are the critical parameters for oriented neurite outgrowth. Anisotropic discontinuous topographies in the form of elliptical microcones at subcellular lengthscale have been strikingly shown to enhance parallel alignment of PNS neurons along the major axis of the features [58]. These results could provide a new research dimension towards topography-controlled nerve cell responses.

Random topographies have been additionally developed, as an approach to simulate the random nanoroughness of the ECM macromolecules and to study the relation between nanotopography alterations imposed by macromolecules and neuronal functions at specific pathological states (e.g. the amyloid plaque buildup in Alzheimer's disease) [35,97,99]. Recent studies show that both hippocampal neurons and PC12 cells seem to selectively sense a specific range of nanoroughness [35]. However, the establishment of a link between the nanoroughness of the substrate and the neural processes remains quite premature. In this context, future studies on hierarchical topographies combining random nanoroughness patterns with deterministic topographies like the grooved topographies are highly desirable. Studies with such platforms could provide a new insight into the topography-induced nerve cell responses.

Although an insight into the mechanism of topography sensing is still lacking, some first conclusions can be drawn based on recent findings [35,56,80,89,96,111]. The topography-driven neuronal polarity on grooved substrates has been associated with ROCK and myosin II- mediated contractility. Studies with micron-sized pillars suggest that neuronal polarization and growth may depend on different mechanisms [111]. Cytoskeletal actin dynamics but not via the Rho/ROCK pathway seem to be involved in pitch-dependent neurite outgrowth in a discontinuous topographical model [96]. Perpendicular neurite contact guidance, which was reported in hippocampal and sensory neurons on continuous anisotropic topographies, seems to have different mechanism from that of the parallel contact guidance [56,80]. In the case of sensory neurons, myosin II seem to be involved in the mechanism of perpendicular contact guidance [80]. While, random nanoroughness has been shown to modulate hippocampal neuronal function via integrin-activated triggering of mechanosensitive ion channels [35].

Most of the studies presented report also on the sensitivity of the neurons to the topographical cues, which can be evident in any interface between a smooth and a topographical patterned substrate, where neurons change its shape, configuration, branching etc. Although it is still challenging to compare the results among the different topographies, it seems that the continuous anisotropic topographies can support a stronger axonal guidance compared with the discontinuous ones of the same lengthscale [90]. Perhaps an interesting case to be explored is the discontinuous anisotropic topographies which have shown a strong topography-induced neuron orientation but this remains to be verified by further studies.

## 4.2 Effect on neuroglial cells

Continuous anisotropic features, including (i) grooves and (ii) fibers, can guide the Schwann cell alignment, migration and influence their functionality.

Schwann cells can orient themselves and align on grooved substrates, although the number of studies remains limited compared with these with neuronal cells [60,114–116]. The minimal groove depth required to provide a guidance effect was found to be 1.5 μm. Neuroglial cells of CNS, including astrocytes and oligodendrocytes tend to sense much shallower grooved features, compared to Schwann cells and also to their neuronal counterparts, such as the hippocampal neurons. Specifically, oligodendrocytes and astrocytes align along grooves down to 100 and 250 nm depth, respectively [67,117]. At the same time, feature width is also a critical parameter for orientation and alignment of the neuroglial cells. Indeed, pattern widths or inter-feature spacing ranging from 2 to 30 μm were found to be optimal for the alignment of Schwann cells. It should be noted that, topography-induced alignment can be strongly enhanced by an intermediate protein coating, for example laminin [115].

Schwann cells on parallel aligned fibers of subcellular size change the shape and orientation state they exhibit on the smooth films and become oriented and aligned along the fibers axis [59,62,63,81]. Interestingly, Schwann cells on larger fibers may encircle the fibers forming chains [81]. While, in whole explant models (e.g. DRG), the topographical anisotropy of the parallel fibers influences the direction of cell migration [48,59,69,70]. Such strong topography-induced guidance of both migrating Schwann cells and outgrowing axons has been investigated in the design of polymeric scaffolds for nerve tissue regeneration after peripheral nerve injury. Remarkably, this topography-guided orientation can be accompanied by changes in maturation and functionality of the Schwann cells [64].

Although the study of the topography-induced guidance effect on neuroglial cells with discontinuous topographies remains very limited, neuroglial cells seem to sense discontinuous features of subcellular lengthscale [58,119]. In particular, parallel oriented elliptical microcones with interfeature distance at the micron scale (i.e. less than 10 μm) promoted oriented Schwann cell growth along the major axis of the ellipse, suggesting that even a discontinuous topographical pattern can promote Schwann cell alignment, provided that it displays a periodic distribution of anisotropic features in a parallel orientation [58].

Recent studies report that random topographies of specific roughness value can affect neuroglial cells (i.e. astrocytes) adhesion and spreading inducing the dissociation of the astrocytes from the neurons [34,35]. Such response has been correlated either with the topography-induced alteration of the astrocyte morphology [35] or the topography-induced selection of the neurons against the astrocytes due to their different morphologies [34]. These studies provide an insight into neurons-neuroglial interactions which can be exploited for the development of proper electrode coatings in neural prosthetic applications where a glial scar tissue formation hinders the neuron-electrode coupling.

Finally, it should be pointed out that although there is an increasing number of studies investigating the topography-sensing mechanism of nerve cells by either studying the cytoskeleton elements or by studying the indirect effect on signaling, similar studies with neuroglial cells are still lacking.

*4.3 Other Issues*

Except for the topographical parameters that have been presented in the previous sections, additional parameters which were beyond the scope of the present review can influence cell-biomaterial interactions.

Material parameters can have a strong impact on the cell responses. For instance, the material type selected determines its physico-mechanical characteristics, for example its stiffness. Indeed, an increasing number of studies reported on the mechanical regulation of cell functions by the substrate elasticity [3,50,120]. Another aspect that has to be taken into consideration is the interdependency between topography and physico-mechanical properties of the substrate. Importantly, the elasticity of the substrate can be controlled geometrically by changing the aspect ratio (geometry-induced elasticity) [94,121], an effect which becomes significant at nanoscale topographical or of high aspect ratio features. Prominent examples are surfaces comprising micro-to-nanoscale post arrays which have been used to study cellular traction forces [122]. Future studies shall systematically study the effect of combination of the topographical parameters with the physico-mechanical ones on cell shape and functions [123].

Topographical models of both non-degradable, (e.g. PDMS, PMMA, etc.) and degradable (i.e. PLGA) polymeric substrates have been presented in this review article. In the case of the degradable substrates, degradation is an important aspect that must be emphasized reflecting the complexity of cell-biomaterials interactions. It is of critical importance to understand both the potential impact of the degradation of the polymers (e.g. type of by-products, rate of degradation, etc.) to the cells and the effect of the cellular environment to the degradation [124,125]. In this context, future studies to evaluate and compare the degradation behaviour of the substrates - in terms of the change of mass, molecular weight and other properties- are highly envisaged [126].

Finally, an important aspect in cell-biomaterials interactions, is protein adsorption. By the time cells reach the surface, the material has already been coated with a monolayer of proteins adsorbed by the supernatant. Hence, the host cells do not actually interface the material, but instead a dynamic layer of proteins [24]. Surface topography at the nanoscale, in the form of either deterministic or random topographies, has a great impact on protein adsorption, in terms of the speciation, conformation and orientation of the surface-bound proteins (reviewed in [127]). The orientation and conformation can have a strong impact on the cell recognition via the integrins. In this context, a future design based on surface nanotopography should focus on the optimization of the protein adsorption [127].

*In conclusion*

New types of cell culture platforms with various topographical patterns at the micro and the nanoscale have been used to study *in vitro* the effect of surface topography on nerve, neuroglial and neural stem cells.

• Continuous topographies in the form of (i) alternating grooves/ridges of subcellular and cellular width and (ii) parallel aligned fibers of subcellular diameters have been shown to strongly enhance guidance and the orientation of neuronal and neuroglial cells along groove and fiber axis, respectively.

• Discontinuous isotropic or anisotropic topographies in the form of pillars, posts and cones, with diameters and inter-feature spacing at subcellular to cellular scale, have been shown to orient neurons and neuroglial cells.

• Random topographies with nano-sized features have been shown to influence specific functions of neurons and astrocytes, as well as their interactions.

Regarding the underlying mechanisms of topography sensing, studies have either focused on the focal adhesion and cytoskeleton elements, or on the indirect effect on cell signaling (for instance, Rho-mediated pathways). A broader insight into such mechanisms is highly envisaged from the future studies. This knowledge will aid in controlling the outgrowth and patterning of nerve and neuroglial cells which is important for a broad spectrum of neuroscience subfields, ranging from basic neurobiological research to tissue engineering.

## Acknowledgements

This work was supported by the European Research Infrastructure NFFA-Europe, funded by EU's H2020 framework programme for research and innovation under grant agreement n. 654360. The help of Zografia Karekou in the design of Figure 1 is acknowledged.

# Controlling the morphology and outgrowth of nerve and neuroglial cells: The effect of surface topography

C. Simitzi[*], A. Ranella and E. Stratakis[*]

*Institute of Electronic Structure and Laser (IESL), Foundation for Research and Technology-Hellas (FORTH), Heraklion, 71003, Greece*

## Statement of significance

There is increasing evidence that physical cues, such as topography, can have a significant impact on the neural cell functions. With the aid of micro-and nanofabrication techniques, new types of cell culture platforms are developed and the effect of surface topography on the cells has been studied. The present review article aims at reviewing the existing body of literature reporting on the use of various topographies to study and control the morphology and functions of cells from nervous tissue, i.e. the neuronal and the neuroglial cells. The cell responses–from phenomenology to investigation of the underlying mechanisms- on the different topographies, including both deterministic and random ones, are summarized.

---

[*] hara.simitzi@gmail.com; stratak@iesl.forth.gr

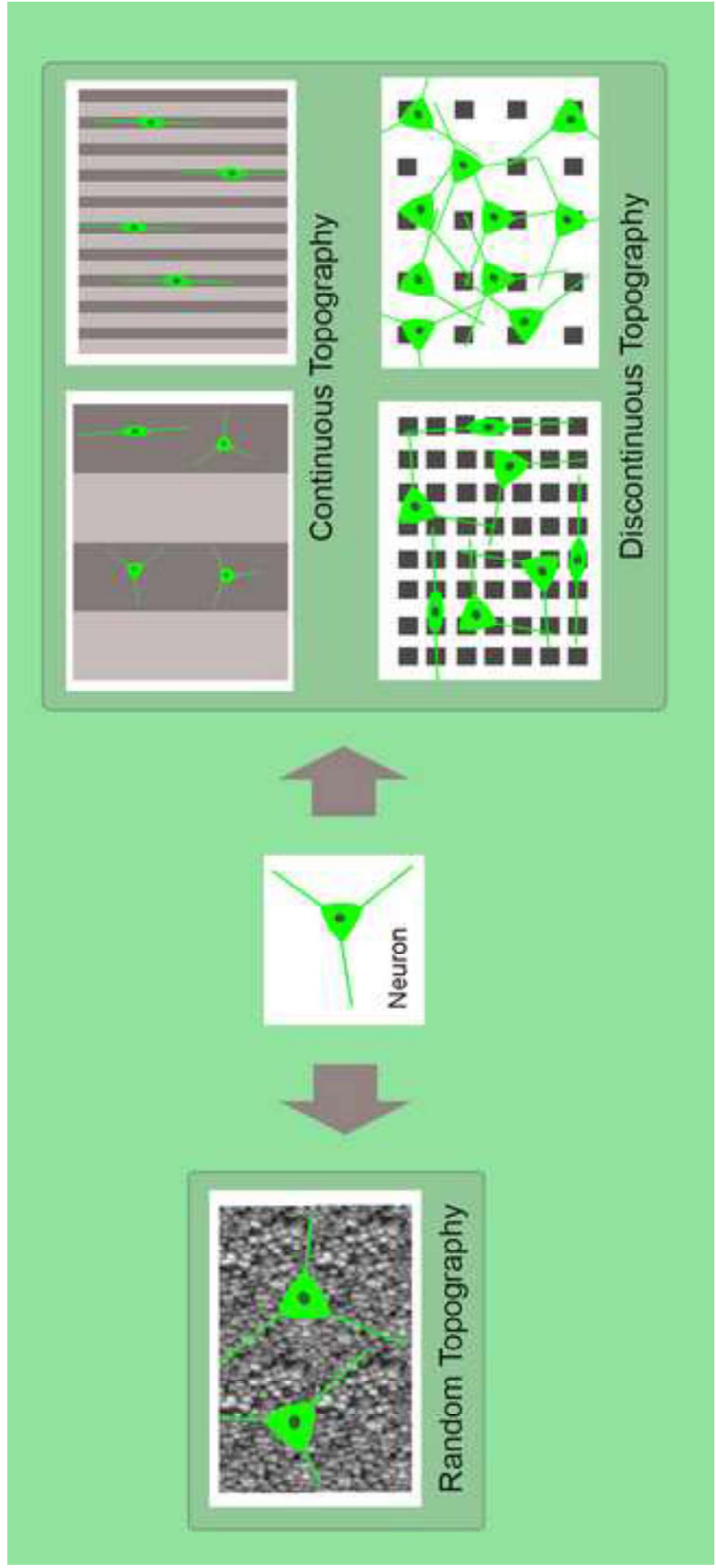

(A) Continuous Topographies

Anisotropic          Isotropic

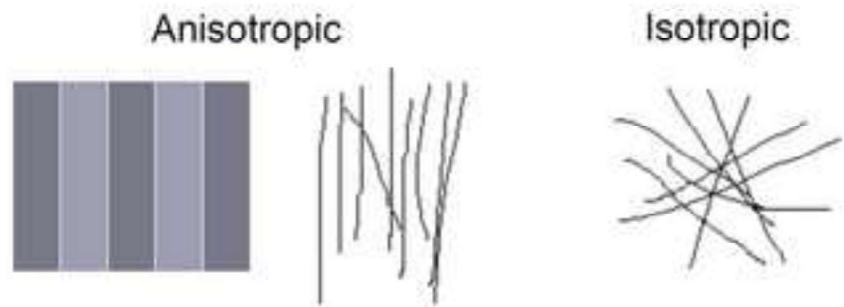

Discontinuous Topographies

Anisotropic          Isotropic

Periodic

Random

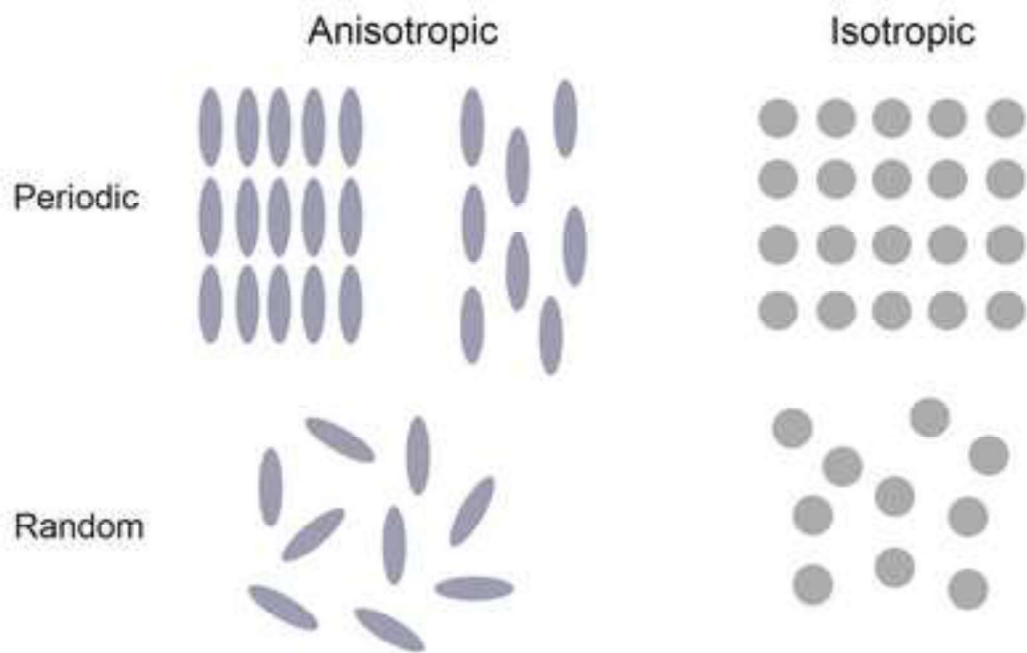

(B) Uniform          Graded

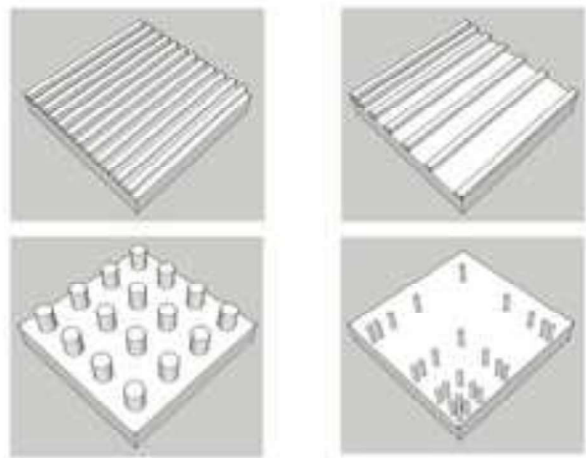

(A)

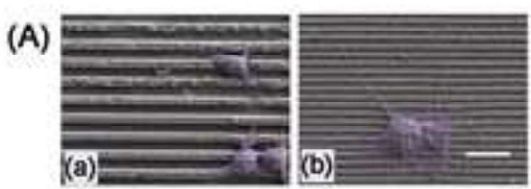

(B)

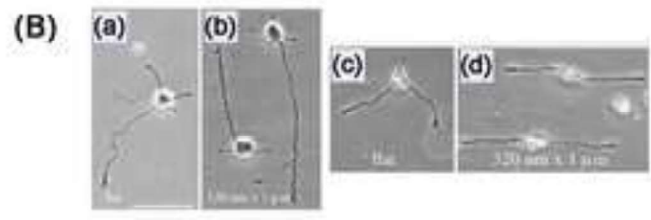

(c)

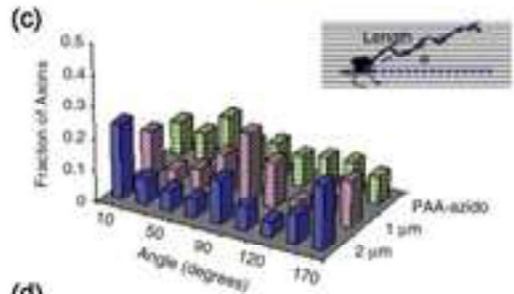

(d)

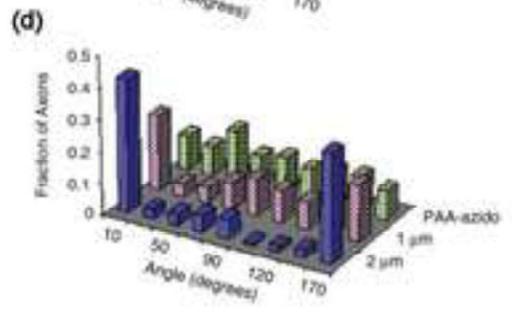

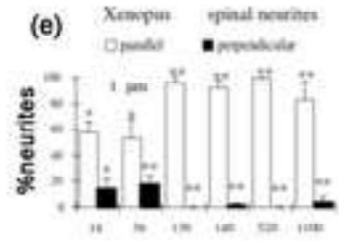

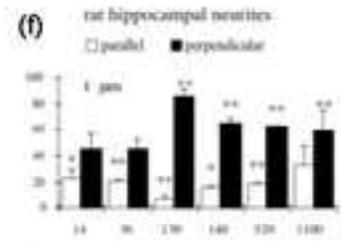

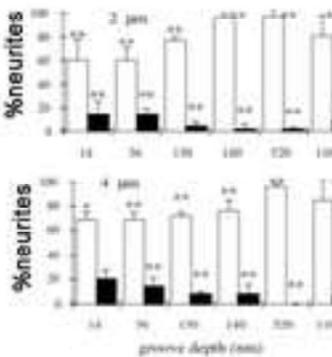

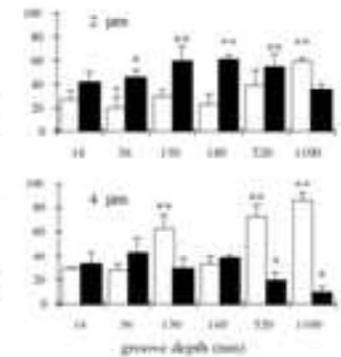

(C)

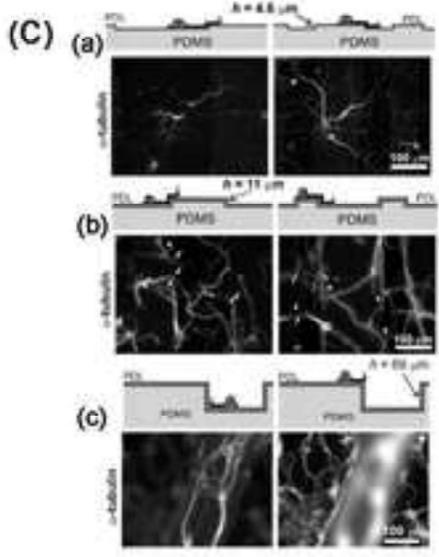

(d)

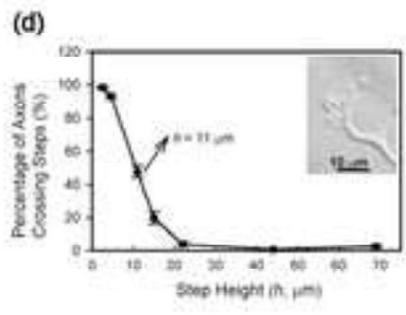

(e)

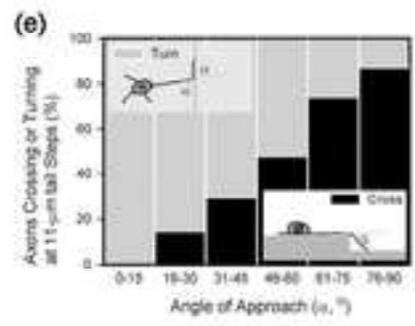

(A) (a) (b) (c) (d) (e) (f)

(g)

| | $L_{Neurite}$ (μm) | $N_{Neurite}$ (# per cell) | $\theta_{Parallel}$ (% neurites) | $\theta_{Perpendicular}$ (% neurites) |
|---|---|---|---|---|
| Control | 12 ± 0.6 | 3.4 ± 0.2 | 32 ± 4 | 33 ± 4 |
| 20 μm | 26 ± 4 | 1.7 ± 0.1 | 62 ± 5 | 11 ± 5 |
| 30 μm | 19 ± 2 | 2.2 ± 0.2 | 58 ± 5 | 12 ± 3 |
| 40 μm | 22 ± 2 | 2.2 ± 0.2 | 59 ± 5 | 17 ± 2 |
| 50 μm | 23 ± 2 | 2.3 ± 0.2 | 50 ± 3 | 20 ± 2 |
| 60 μm | 21 ± 2 | 2.7 ± 0.2 | 44 ± 2 | 14 ± 2 |

(B) (a) (b) (c)

(C)

| | BF | Tau-1 | Merge |
|---|---|---|---|
| a | A3 | | |
| b | F2 | | |
| c | B3 | | |

(d) Axon Length (μm) — Gratings, Circles, Dots, Flat

(e) Dendrite Branches/ Cell Body Pixels — Gratings, Circles, Dots, Flat

(D) (a) (b) Number of bridges vs. groove width (μm) (c) Number of bridges vs. plateau width (μm)

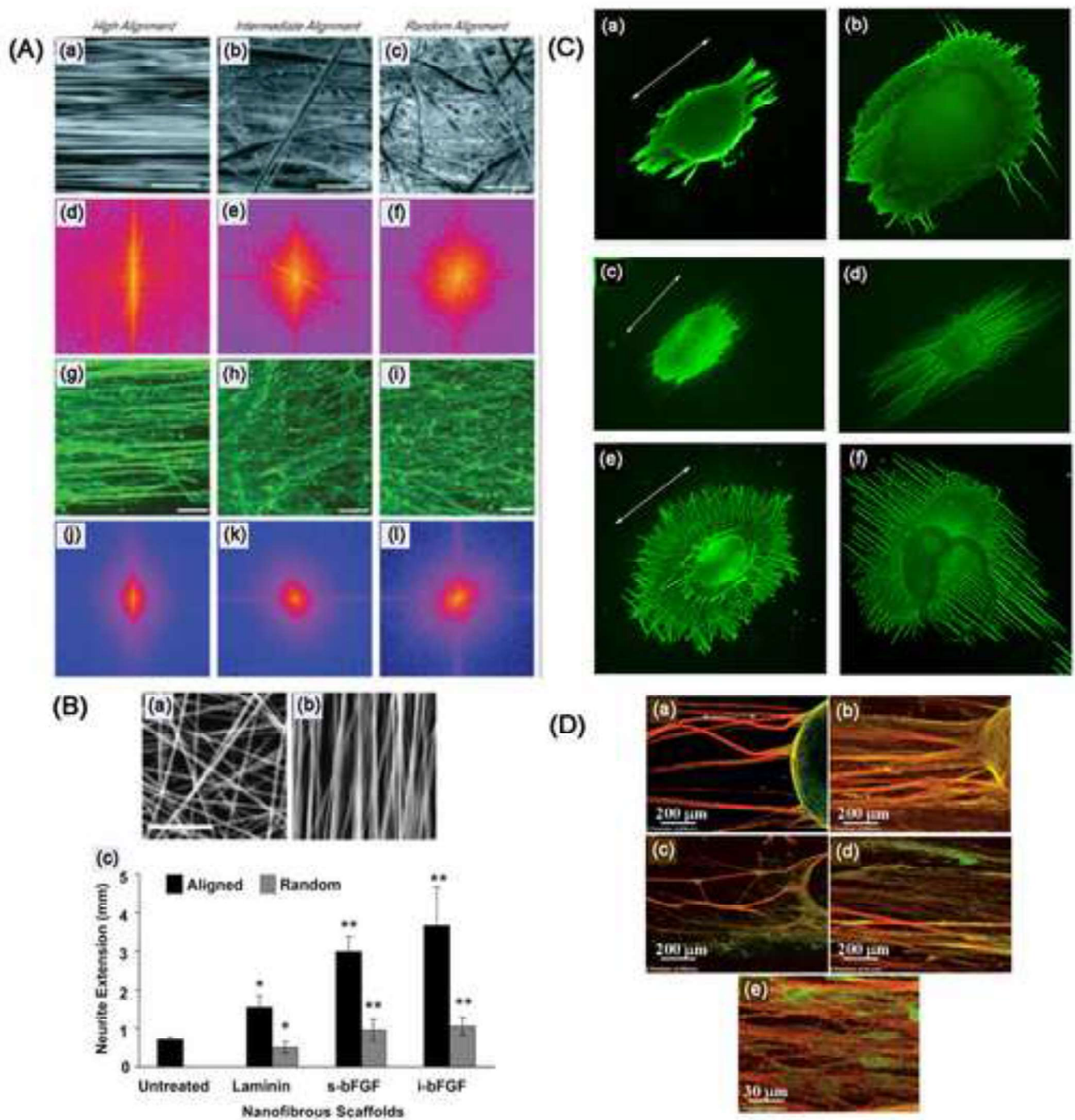

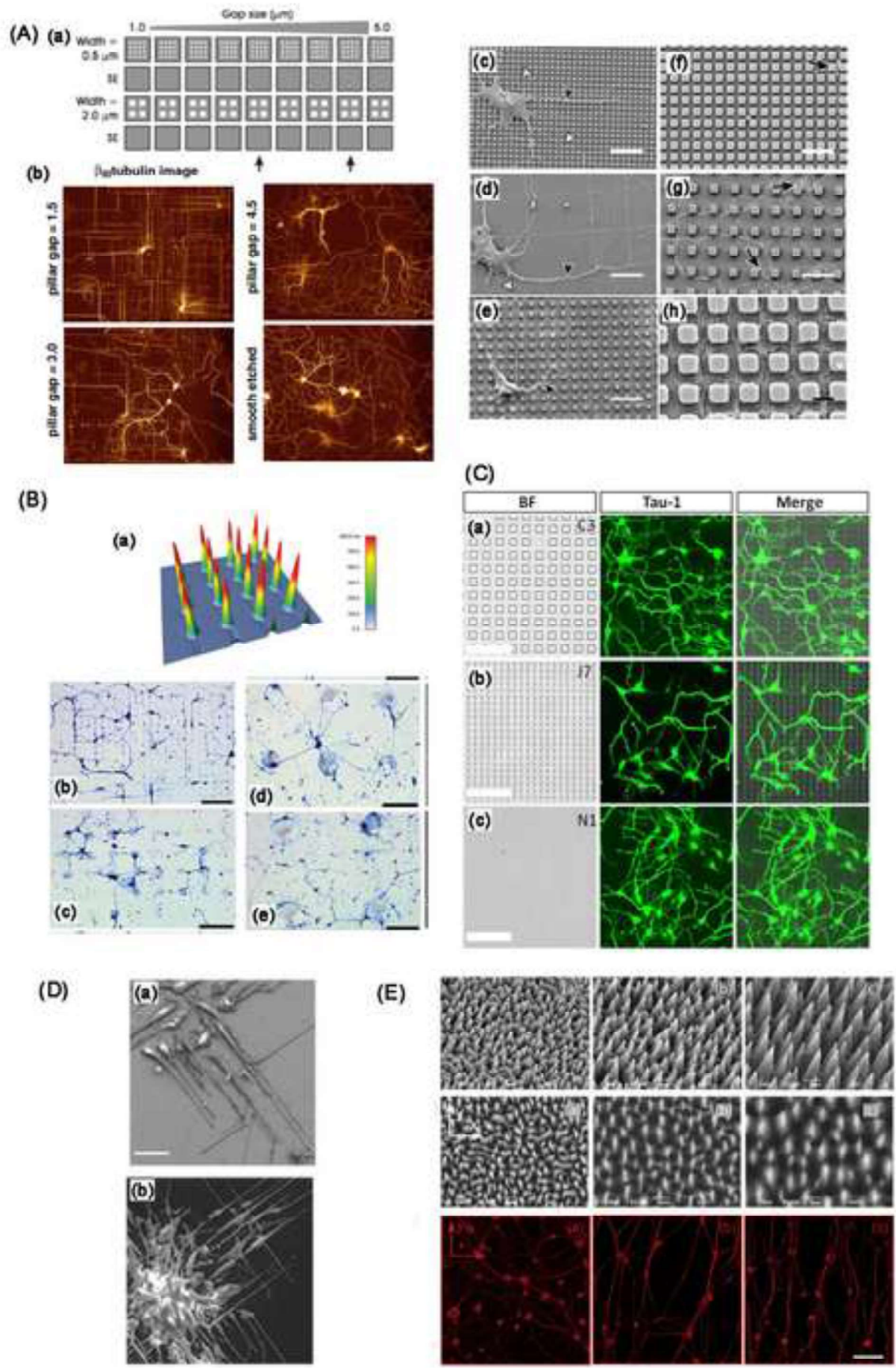

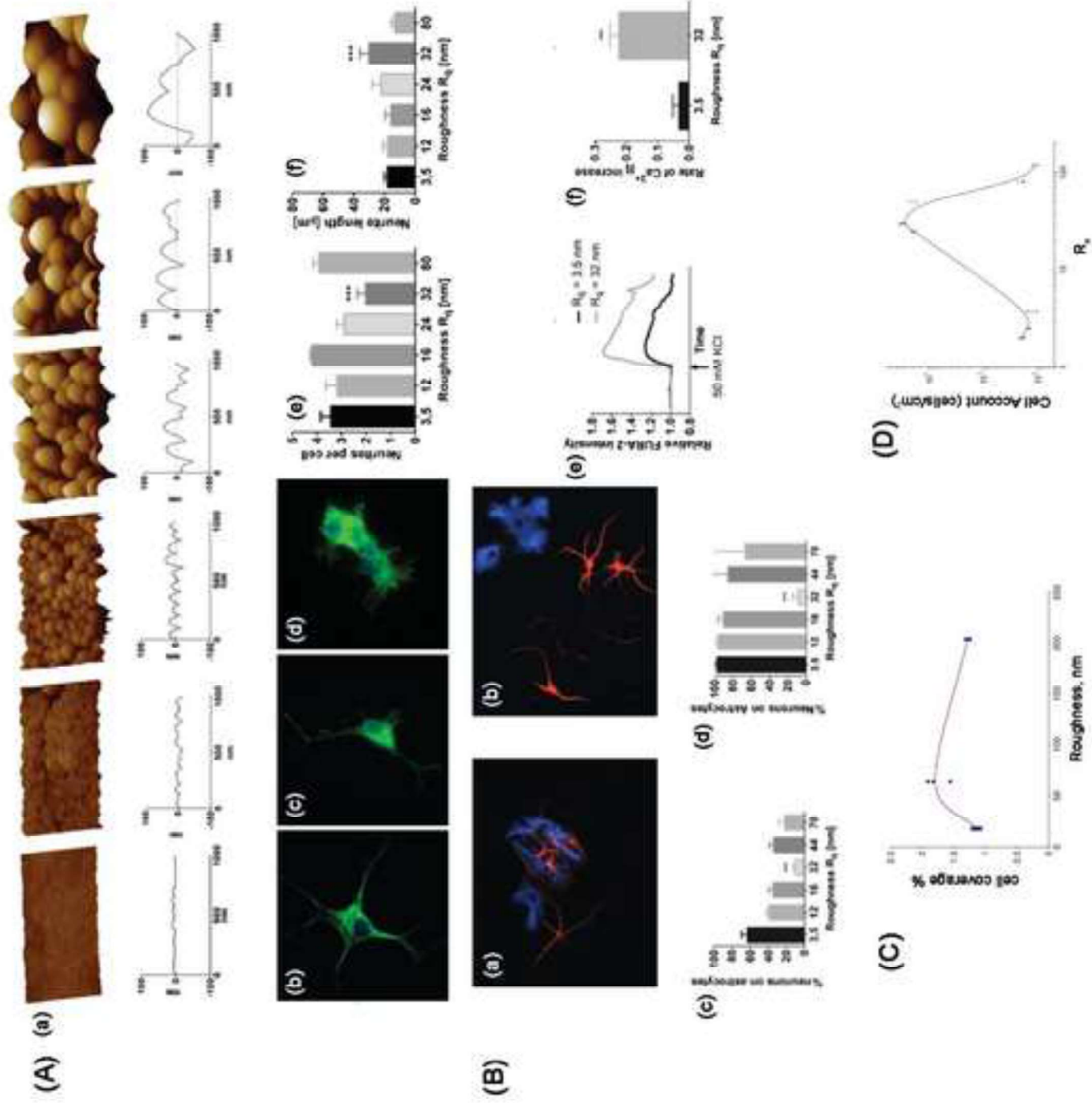

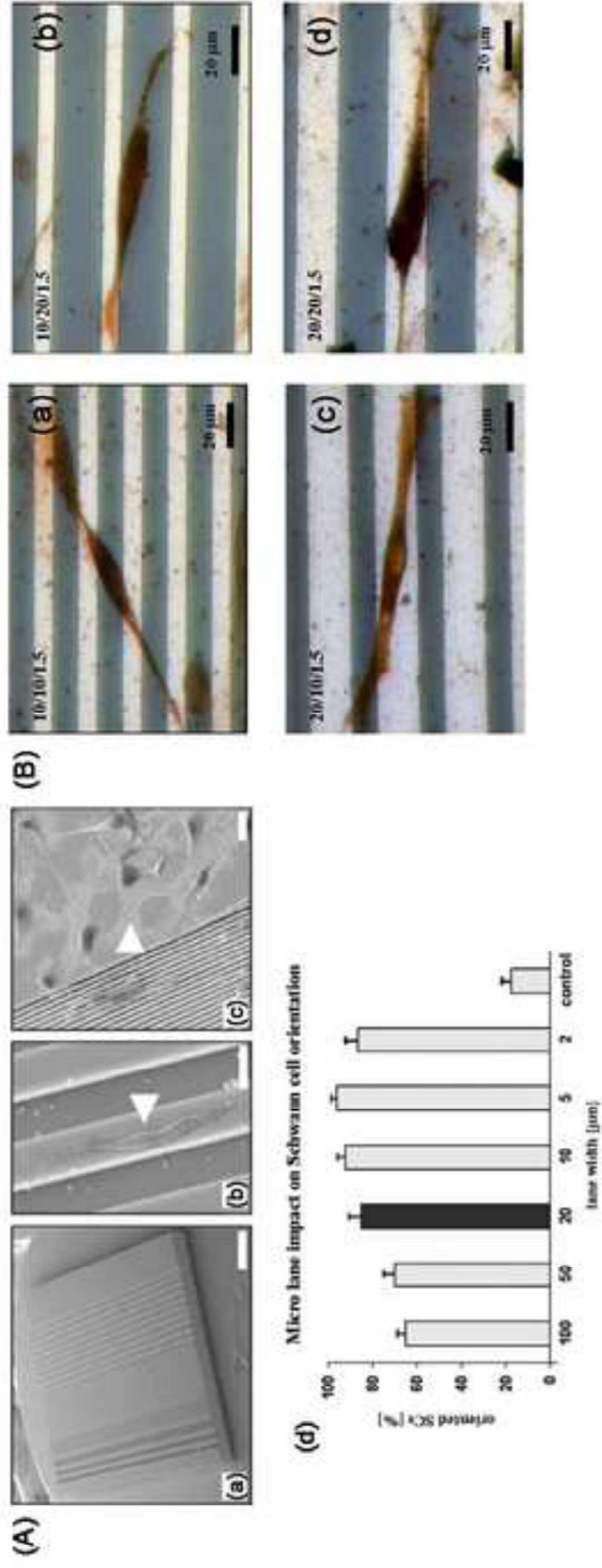

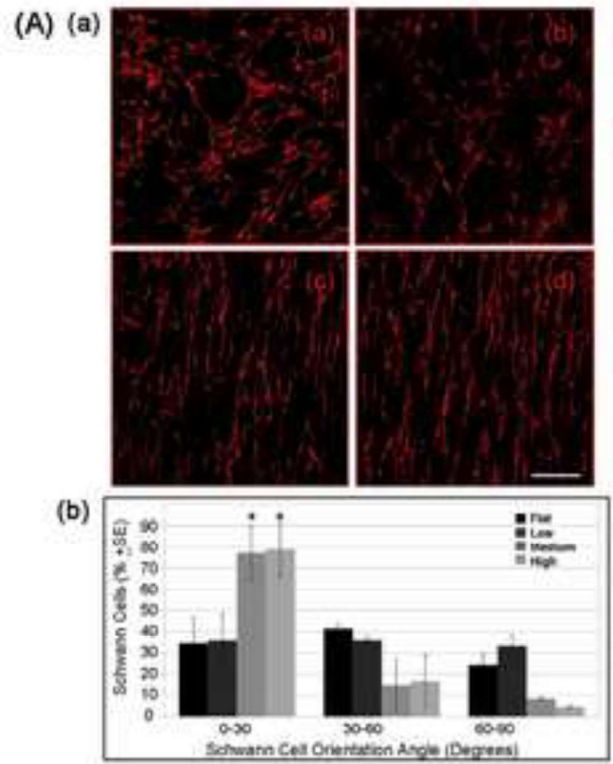

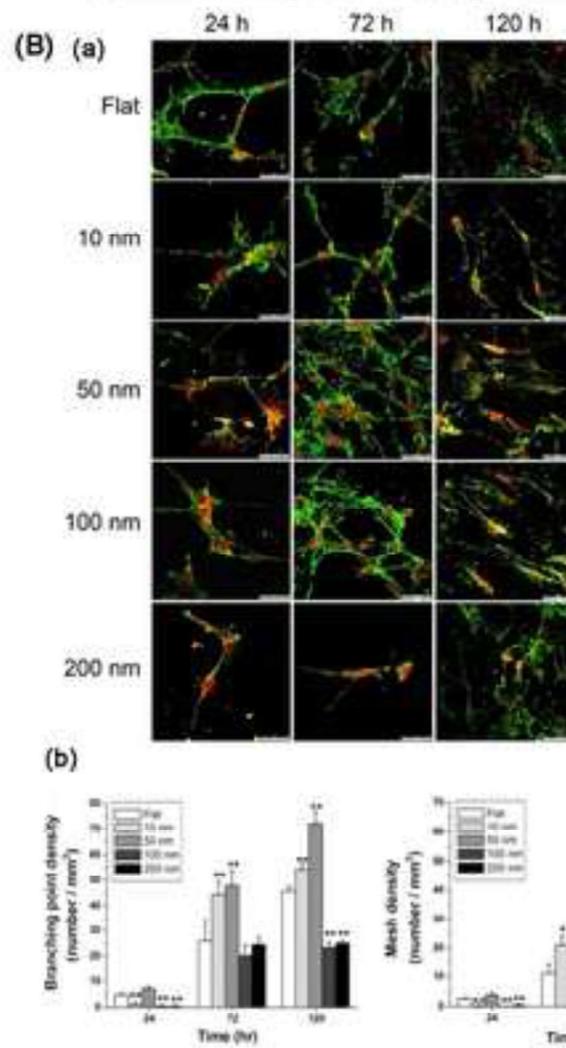

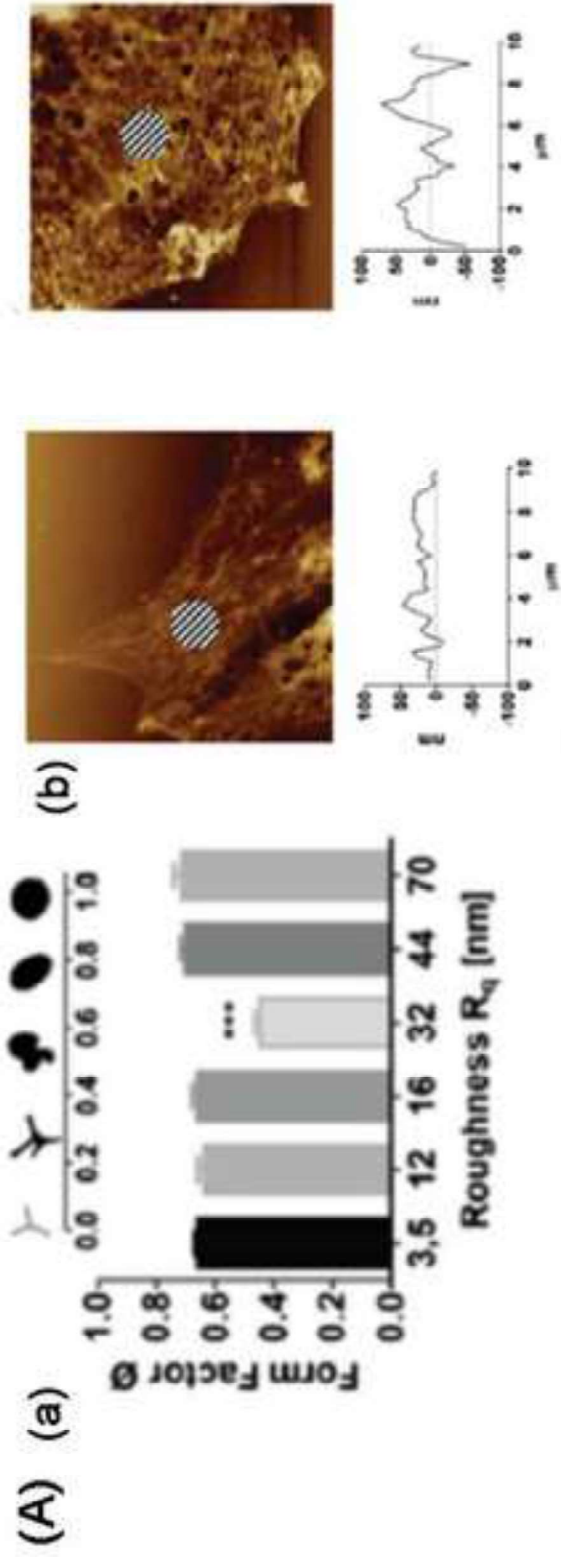

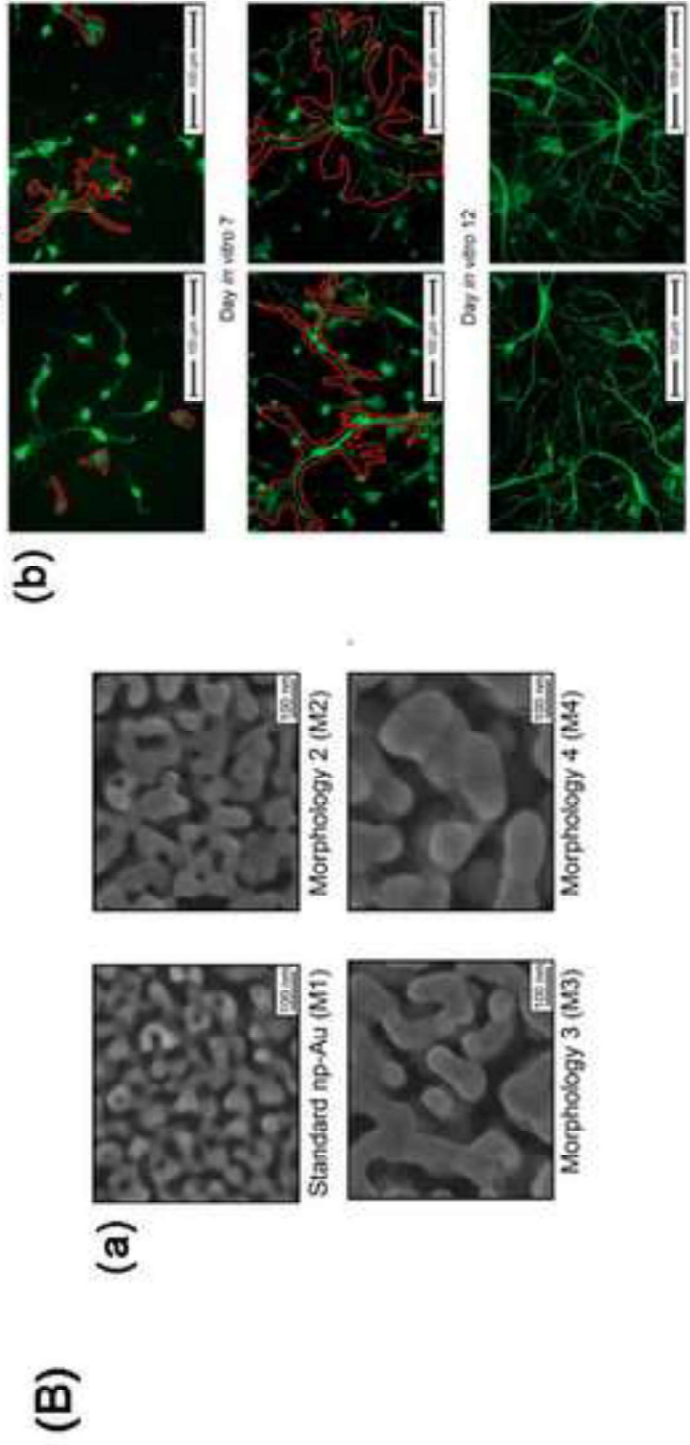

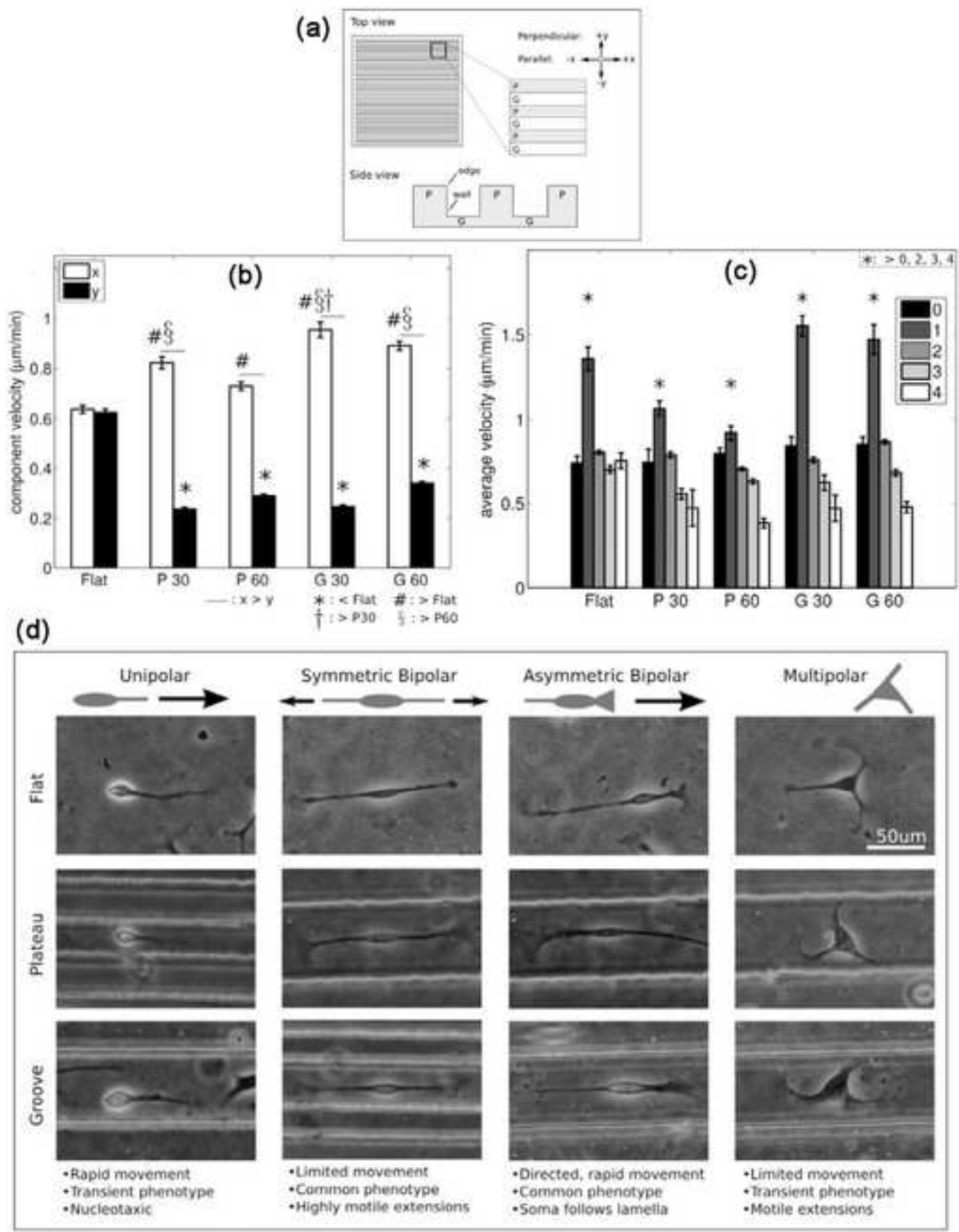